\documentclass[traditabstract]{aa} 
\usepackage{graphicx}
\usepackage{txfonts}
\usepackage{amssymb}
\usepackage{array}
\usepackage{dcolumn}
\usepackage[round]{natbib}
\bibpunct{(}{)}{;}{a}{}{,}

\begin{document}
\title{Nebular and global properties of the gravitationally lensed galaxy
``the 8~o'clock arc''
\thanks{Based on X-shooter observations made with the European Southern
Observatory VLT/Kueyen telescope, Paranal, Chile, collected under the programme 
ID No.~284.A--5006(A).}$^,$
\thanks{Based on observations made with the NASA/ESA Hubble Space Telescope, 
obtained from the data archive at the Space Telescope Science Institute.}}


\author{M. Dessauges-Zavadsky\inst{1}, 
	L. Christensen\inst{2},
	S. D'Odorico\inst{3},
	D. Schaerer\inst{1,4},
	\and
	J. Richard\inst{5,6}
        }

\offprints{miroslava.dessauges@unige.ch}

\institute{Observatoire de Gen\`eve, Universit\'e de Gen\`eve, 51 Ch. des
           Maillettes, 1290 Sauverny, Switzerland
           \and
	   Excellence Cluster Universe, Technische Universit\"at M\"unchen,
	   Bolzmanstrasse 2, 85748 Garching, Germany
	   \and
           European Southern Observatory, Karl-Schwarzschildstrasse 2, 85748
           Garching, Germany
	   \and
	   Laboratoire d'Astrophysique de Toulouse-Tarbes, Universit\'e de
           Toulouse, CNRS, 14 Avenue E. Belin, 31400 Toulouse, France
	   \and
	   Dark Cosmology Center, Niels Bohr Institute, University of
	   Copenhagen, Juliane Maries Vej 30, 2100 Copenhagen, Denmark
	   \and
	   Centre de Recherche Astrophysique de Lyon, Universit\'e Lyon 1, 
	   Observatoire de Lyon, 9 Avenue Charles Andr\'e, 69561 Saint Genis 
	   Laval cedex, France
           }

\date{}

\authorrunning{Dessauges-Zavadsky et~al.}

\titlerunning{Rest-frame optical spectrum of the 8~o'clock arc}

 
\abstract{We present the analysis of new near-infrared, intermediate-resolution 
spectra of the gravitationally lensed galaxy ``the 8~o'clock arc'' at 
$z_{\rm sys} = 2.7350$ obtained with the X-shooter spectrograph on the Very 
Large Telescope. These rest-frame optical data, combined with Hubble and 
Spitzer Space Telescopes images, provide very valuable information, which 
nicely complement our previous detailed rest-frame UV spectral analysis, and 
make the 8~o'clock arc one of the better understood ``normal'' star-forming 
galaxies at this early epoch of the history of the Universe. From 
high-resolution HST images, we reconstruct the morphology of the arc in the 
source plane, and identify that the source is formed of two majors parts, the 
main galaxy component and a smaller blob separated by 1.2\,kpc in projected 
distance. The blob, with a twice larger magnification factor, is resolved in 
the X-shooter spectra. The multi-Gaussian fitting of detected nebular emission 
lines and the spectral energy distribution modeling of the available 
multi-wavelength photometry provide the census of gaseous and stellar dust 
extinctions, gas-phase metallicities, star-formation rates (SFRs), and stellar, 
gas, and dynamical masses for both the main galaxy and the blob. As a result, 
the 8~o'clock arc shows a marginal trend for a more attenuated ionized gas than 
stars, and supports a dependence of the dust properties on the SFR. With a high 
specific star-formation rate, $\rm SSFR = 33\pm 19\,Gyr^{-1}$, this lensed 
Lyman-break galaxy deviates from the mass-SFR relation, and is characterized by 
a young age of $40^{+25}_{-20}\,\rm Myr$ and a high gas fraction of about 72\%. 
The 8~o'clock arc satisfies the fundamental mass, SFR, and metallicity 
relation, and favors that it holds up beyond $z\simeq 2.5$. We believe that the 
blob, with a gas mass $M_{\rm gas} = (2.2\pm 0.9)\times 10^9\,\rm M_{\sun}$ 
(one order of magnitude lower than the mass of the galaxy), a half-light radius 
$r_{1/2} = 0.53\pm 0.05\,\rm kpc$, a star-formation rate $\rm SFR_{H\alpha} = 
33\pm 19\,M_{\sun}\,yr^{-1}$, and in rotation around the main core of the 
galaxy, is one of these star-forming clumps commonly observed in $z>1$ 
star-forming galaxies, because it is characterized by very similar physical 
properties. The knowledge of detailed physical properties of these clumps is a 
very useful input to models that aim to predict the formation and evolution of
these clumps within high-redshift objects.}
 

\maketitle

%

\section{Introduction}

With the advent of the 8--10\,m class optical telescopes on the ground and of 
the latest generation of space observatories working from X-rays to 
far-infrared (IR) wavelengths, it has become possible to carry out 
multi-wavelength spectroscopy and imaging of thousands of galaxies at redshifts 
between 2 and 4, namely at the peak of the star-formation history of the 
Universe. In particular, star-forming galaxies, known also as Lyman-break 
galaxies (LBGs), easily identified by a break in their ultraviolet (UV) 
continuum that is caused by the Lyman limit from intergalactic and interstellar 
\ion{H}{i} absorption below 912\,\AA, have been the targets of most studies 
\citep{steidel96}. At $z>2$ their absorption from the intergalactic medium 
(IGM) is more pronounced and the galaxy UV flux is redshifted toward optical 
wavelengths, where the ground-based telescopes and their instruments have their 
maximum detection efficiency, and the blocking effect of the atmosphere and the 
emission from the night sky have its minimum. Color and low-resolution spectral 
information has been gathered for many of these generally faint galaxies 
($R\simeq 24.5$ at $z\sim 3$), whereas very long integration times are 
required, even at the larger telescopes, to acquire intermediate-resolution 
spectra, which are yet needed to derive their detailed individual physical 
properties. To gain insights on their average properties, one is thus mostly 
forced to rely on stacked low-resolution spectra \citep{shapley03,vanzella09}.

Instead of waiting for the next-generation telescopes with large collecting 
areas to obtain good resolution, good signal-to-noise ratio spectra of LBGs, we 
can take advantage of the light magnification provided by gravitational 
lensing. In the best cases, the background galaxies can benefit of a boost of 
their total flux by a factor of $30-50$. This implies that 
intermediate-resolution spectroscopy of individual galaxies, which are 
intrinsically $3-4$ magnitudes fainter than they appear at the telescope, 
becomes achievable with current instrumentation. As an important byproduct, 
this approach gives access to a more representative range of the luminosity 
function of galaxies at $z\sim 2-4$. 

A handful of studies of these highly magnified star-forming galaxies yielded
the physical properties of their stellar population and their interstellar 
medium (ISM), as well as their dynamical properties. \citet{pettini02} 
pioneered this study approach with the extraordinary bright LBG MS\,1512-cB58
\citep[see also][]{pettini00,teplitz00,savaglio02}. In the last decade, new
search techniques applied to the Sloan Digital Sky Survey (SDSS) and Hubble 
Space Telescope (HST) images led to the identification of new strongly-lensed, 
high-redshift galaxies. The brightest galaxies were already targeted for 
detailed studies at rest-frame UV and/or optical wavelengths \citep{lemoine03,
swinbank07,swinbank09,stark08,cabanac08,hainline09,quider09,quider10,pettini10,
christensen10,jones10,bian10,rigby11,richard11}. In the era where one of the 
major objectives of astrophysics is to determine the precise physical 
properties of high-redshift, star-forming galaxies, the analysis of 
strongly-lensed LBGs provide very complementary results and informations to the 
huge efforts done with galaxies selected in blank fields 
\citep[e.g.,][]{erb06a,erb06b,erb06c,law07,maiolino08,genzel08,mannucci09,
forster06,forster09}.

Among the highly magnified objects, a particularly interesting target is the
``8~o'clock arc'', discovered by \citet{allam07}. The lensing by the $z=0.38$ 
luminous red galaxy SDSS\,J002240.91+143110.4 distorts this Lyman-break galaxy 
at $z_{\rm sys} = 2.7350$ into four separate images that correspond to four 
different images of the \textit{same} object. Three of them, labeled A1, A2, 
and A3, form a partial Einstein ring of radius $\theta_{\rm E} = 3.32\arcsec 
\pm 0.16\arcsec$, extending over $9.6\arcsec$ in length (see 
Fig.~\ref{fig:reconstruction}, left-hand panel). \citet{finkelstein09} carried 
out a comprehensive study of the 8~o'clock arc with the help of low-resolution 
spectra from UV to the $K$ band. In a first paper \citep[][hereafter DZ10]
{dessauges10} we used intermediate-resolution spectra obtained with the newly 
installed X-shooter spectrograph \citep{dodorico06} that cover the observed 
range from 320 to 1000\,nm to further improve the knowledge of properties of 
this lensed LBG. We derived its stellar and ISM metallicities for the first 
time and highlighted the main ISM line similarities and differences observed 
among the few lensed LBGs studied in detail so far. The high quality of the 
data allowed us also to model the Ly$\alpha$ line profile with 3D radiation 
transfer codes, and to obtain results fully consistent with the scenario 
proposed earlier, in which the diversity of Ly$\alpha$ line profiles in LBGs 
and Ly$\alpha$ emitters, from absorption to emission, can mostly be explained 
by \ion{H}{i} column density and dust content variations 
\citep{verhamme06,verhamme08}.

In this paper we use new X-shooter observations, which now include the
near-infrared spectral range and cover the nebular emission lines of the 
8~o'clock arc, and archive HST and Spitzer imaging data to complete our 
detailed picture of this unique galaxy. The new physical and morphological 
properties derived make the 8~o'clock arc one of the most deeply studied 
``normal'' star-forming galaxies at this early epoch of the history of the 
Universe. In Sect.~\ref{sect:obs+reduction} we present the spectroscopic and 
imaging observations, and the corresponding data reduction procedures. In 
Sect.~\ref{sect:analysis} we deal with the photometry and the analysis of 
spectra, in particular we provide the gravitational lens model of the arc and 
the fits of the nebular emission lines with multi-Gaussian profiles. In 
Sect.~\ref{sect:properties} we revisit the properties of the galaxy (dust
extinction, metallicity, star-formation rate, age, and stellar, gas, and 
dynamical masses) as derived from the combination of spectroscopic and imaging 
data. A final summary of the results and their discussion is presented in 
Sect.~\ref{sect:discussion}. Throughout the paper, we assume a $\Lambda$-CDM 
cosmology with $\Omega_{\Lambda} = 0.73$, $\Omega_{\rm M} = 0.27$ and 
$h = 0.71$. All magnitudes are given in the AB system.

%

\section{Observations and data reduction}
\label{sect:obs+reduction}

\subsection{X-shooter spectroscopy}


X-shooter is the first of the second-generation instruments on the Very 
Large Telescope (VLT) at Cerro Paranal, Chile, in operation at the European 
Southern Observatory (ESO) since October 2009. It consists of three Echelle 
spectrographs with prism cross-dispersion, mounted on a common structure at the 
Cassegrain focus of the Unit Telescope 2. The light beam from the telescope is 
split by two dichroics that direct the light in the spectral ranges of 
$300-560$\,nm and $560-1015$\,nm to the slit of the ultraviolet-blue (UV-B) and 
visual-red (VIS-R) spectrographs, respectively. The undeviated beam in the 
spectral range of $1025-2400$\,nm feeds the near-infrared (NIR) spectrograph. A 
full description of the instrument is provided by \citet{vernet10}.

A first set of X-shooter observations of the 8~o'clock arc, made solely with 
the UV-B and VIS-R spectrograph arms, was obtained during the first 
commissioning run in November 2008, and allowed a detailed analysis of the 
rest-frame UV spectrum of this lensed LBG (see DZ10). In November and December 
2009, we got additional observations in Director's Discretionary Time (program 
ID No.~284.A--5006(A)) to complete the 8~o'clock arc data set with the NIR 
spectrograph arm and to obtain the rest-frame optical spectrum. The $11\arcsec$ 
long entrance slit was rotated to the same position angles on the sky as for 
the first set of observations, $\rm PA = 121\degr$ and $13\degr$ aligned along 
the 8~o'clock images A2 and A3 and along the image A2 and the galaxy lens, 
respectively (see Fig.~1 in DZ10). A total exposure time of $4\times 1200$\,s 
per position angle was obtained in good conditions, with clear sky, seeing 
$< 1\arcsec$, and airmass $< 1.6$. Slit widths of $1.3\arcsec$ in the UV-B, 
$1.2\arcsec$ in the VIS-R, and $0.9\arcsec$ in the NIR were used, corresponding 
to resolving powers $R\equiv \lambda/\Delta \lambda = 4000$, 6700, and 5600, 
respectively. The observations used a nodding along the slit approach, with a 
typical offset between individual exposures of $5\arcsec$ for $\rm PA = 
121\degr$ and $3\arcsec$ for $\rm PA = 13\degr$, adapted to avoid any 
overlapping in the combination of consecutive exposures.

The data were reduced with a preliminary version of the ESO X-shooter pipeline
\citep{goldoni06,modigliani10}. Bad pixels were found using calibration frames, 
and cosmic ray hits were detected and cleaned using the L.A. Cosmic routine
\citep{dokkum01}. In the NIR arm, the sky background emission was subtracted 
with the help of adjacent exposures where the two components, either 
$\rm A2-A3$ or $\rm A2-galaxy$ lens, were offset along the slit. Then, the data 
were flat-fielded, and wavelength-calibrated, and the trace of each order 
detected using calibration frames. The final products from the pipeline are 2D 
rectified spectra with the individual orders merged in a weighted scheme, using 
the error spectra derived and propagated in the pipeline.

The respective 1D spectra of the images A2 and A3 of the 8~o'clock arc were 
extracted from the 2D pipeline-produced spectra, using adapted extraction 
windows. They were then co-added using their signal-to-noise ratios (S/N) as 
weights, when several exposures of the same lensed image observed with the same 
spectrograph arm and at the same position angle were available. Preliminarily 
the 1D NIR spectra were corrected for telluric absorption by dividing them by 
the normalized spectra of the B5V star Hip022840 and the B5 star Hip023946, 
observed with the same instrumental set-up and at approximately the same 
airmass as the 8~o'clock arc\footnote{The flux of the 8~o'clock arc spectra was 
set to zero whenever the atmospheric transmission was below 10\%.}. Absolute 
flux calibration applied to the NIR spectra was based on the standard star 
BD+17\,4708 whose spectrum was recorded during the same nights as the 8~o'clock 
arc. To derive the transmission of X-shooter, we used the fluxes of this star 
measured with the HST as reference \citep{bohlin07}.

%

\begin{figure*}[!]
\centering
\includegraphics[height=7.5cm,clip]{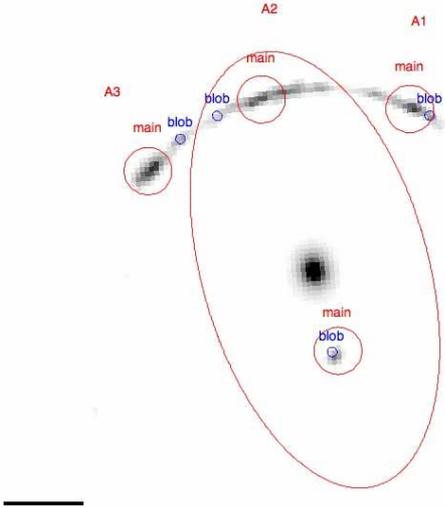}\hspace{1cm}
\includegraphics[height=7.3cm,clip]{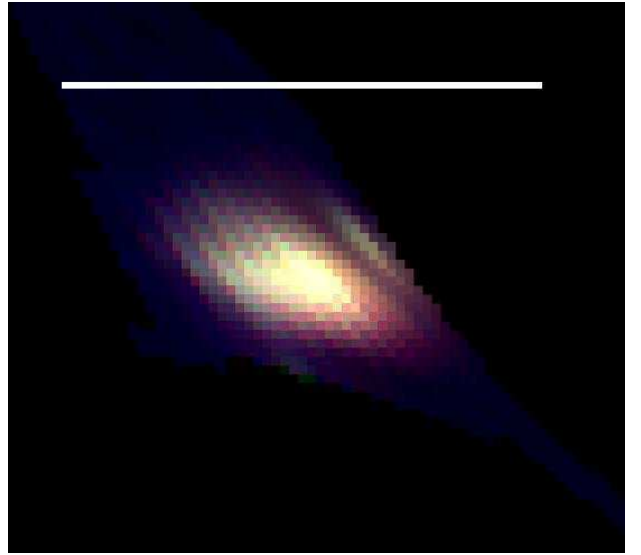}
\caption{{\it Left.} WFPC2/F450W black/white image of the 8~o'clock arc showing 
the positions of the four lensed images of the main galaxy component in red and 
the blob in blue. The two brightest images labeled A2 and A3 were targeted for 
spectroscopy with X-shooter. The large ellipse corresponds to the external 
critical line at the redshift of the arc as derived from lens modeling 
(Sect.~\ref{sect:reconstruction}). {\it Right.} WFPC2/F450W-F650W-F814W 
composite-color image showing the reconstructed morphology of the 8~o'clock arc 
in the source plane, as derived from the best-fit lens model. The source is 
formed of two major parts, the main galaxy component (on the left) and a 
smaller blob (on the right) separated by 0.15\,arcsec, or equivalently 1.2\,kpc 
in projected distance. The blob has a twice higher magnification factor than 
the main galaxy component. In both panels, the horizontal scale bar refers to 
1\,arcsec.}
\label{fig:reconstruction}
\end{figure*}
%

\subsection{HST imaging}


We made use of the high-resolution optical and NIR imaging data taken with the 
Wide Field Planetary Camera 2 (WFPC2) and Near Infrared Camera and Multi-Object 
Spectrometer 2 (NICMOS2) instruments on the Hubble Space Telescope, which are 
available from the HST archive under the program ID No.~11167 (PI: S.~S. 
Allam). The 8~o'clock arc is clearly resolved, and was observed in five bands 
$B$, $V$, $I$, $J$, and $H$, with the respective filters WFPC2/F450W, 
WFPC2/F606W, WFPC2/F814W, NIC2/F110W, and NIC2/F160W. A total exposure time of 
$4\times 1100\,\rm s$ per $BVI$ bands, $5120\,\rm s$ in the $J$ band, and 
$4\times 1280\,\rm s$ in the $H$ band was obtained and consists of independent 
frames. The WFPC2 $BVI$ frames, with a pixel scale of $0.1\arcsec$, were 
arranged in a four-point dither pattern, with random dithered offsets between 
individual exposures within $1\arcsec$ in right ascension and declination. The 
NICMOS2 $JH$ frames, with a pixel scale of $0.075\arcsec$, were also arranged 
in a four-point dither pattern, but with offsets between individual exposures 
of $2.5\arcsec$.

To combine independent WFPC2 frames into a final frame for each band and reject 
cosmic ray hits, we used the drizzle routine of \citet{fruchter02}, which also 
corrects for instrument geometric distortions. In the drizzling, we chose the 
parameters $\rm pixfrac=0.8$ and $\rm scale=1.0$ to be able to effectively
remove noisy pixels from cosmic ray hits, but also extract the morphological
information of the lensed structure without introducing artifacts. The chosen
parameters resulted in a point-spread function (PSF) FWHM of $0.18-0.2\arcsec$ 
for the WFPC2 images. A similar procedure was used for the reduction of 
NICMOS2 frames, but a number of instrument-specific improvements were included 
(such as the flagging of hot/cold pixels), following the prescriptions given in 
\citet{richard08}. This yielded a PSF FWHM of $0.12\arcsec$ for the NICMOS2 
images. All HST images were then registered onto one another, using bright 
isolated point sources in the common fields to ensure precise photometry over 
the same regions of the 8~o'clock arc.

%

\begin{table}
\caption{HST and Spitzer photometry of the 8~o'clock arc images A2 and A3}
\label{tab:photometry}
\centering                
\begin{tabular}{l c | c | c}  
\hline\hline    
       &      & A2           & A3 \\
Filter & Band & AB magnitude & AB magnitude \\
\hline              
F450W  & $B$ & $21.94\pm 0.10$ & $22.04\pm 0.10$ \\
F606W  & $V$ & $21.36\pm 0.10$ & $21.37\pm 0.10$ \\
F814W  & $I$ & $21.10\pm 0.10$ & $21.21\pm 0.10$ \\
F110W & $J$ & $21.12\pm 0.10$ & $21.27\pm 0.10$ \\
F160W & $H$ & $20.77\pm 0.10$ & $20.83\pm 0.10$ \\
$\rm F160W - 3.6\,\mu m$ & IR color & \multicolumn{2}{c}{$0.61\pm 0.12$} \\
$\rm F160W - 4.5\,\mu m$ & IR color & \multicolumn{2}{c}{$0.93\pm 0.12$} \\
$\rm F160W - 5.8\,\mu m$ & IR color & \multicolumn{2}{c}{$0.94\pm 0.12$} \\
$\rm F160W - 8.0\,\mu m$ & IR color & \multicolumn{2}{c}{$0.78\pm 0.12$} \\
\hline                                
\end{tabular}
\tablefoot{
The tabulated AB magnitudes of A2 and A3 correspond to the total photometry of 
the components main+blob as defined in Sect.~\ref{sect:reconstruction}. The 
errors on the magnitudes were derived from the noise measured in the images, 
scaled to the size of the aperture used for color measurements. A 0.05 
magnitude error in the zero-point calibration was added in quadrature.
}
\end{table}
%

\section{Analysis}
\label{sect:analysis}

\subsection{Photometry}

%

Because the 8~o'clock arc is fully resolved in the HST images, we could derive 
the overall photometry in the multiple images of the lensed LBG and, in 
particular of A2 and A3 chosen for X-shooter spectroscopy. The measurements 
were performed using the SExtractor software \citep{bertin96} in the 
``double-image'' mode. We used the WFPC2/F450W band as the detection image, and 
measured the fluxes in $1.0\arcsec$ diameter apertures across all HST bands. 
The WFPC2/F450W band was chosen as the detection image because it offers the 
best contrast to detect the 8~o'clock arc images and to separate their light 
from the lensing luminous red galaxy. $1.0\arcsec$ diameter apertures were 
chosen to match the $0.9\arcsec$ slit width of the X-shooter NIR spectra, which
then allow a direct comparison between HST images and ground based spectra. For 
NICMOS2 photometry, the WFPC2 images had to be first convolved with a Gaussian 
filter matching the slightly different FWHM of the NICMOS2 PSF. This ensured 
accurate color measurements between the WFPC2/F450W and NICMOS2 bands. Then, an 
estimate of the total flux of A2 and A3 was provided by \texttt{MAGAUTO} 
measured in the detection image. We checked from the segmentation image that 
the regions used by SExtractor for the flux measurements correspond well to the 
entire A2 and A3 counterparts in the image. We estimated an error of 
$0.1\,\rm mag$ in the absolute flux normalization, but this does not affect the 
colors used for the SED fitting (see Sect.~\ref{sect:SED}). The final 
photometry of the lensed images A2 and A3, normalized to the absolute flux in 
the detection image, is given in Table~\ref{tab:photometry}.

The 8~o'clock arc was also observed with the Infrared Array Camera (IRAC) on 
the Spitzer Space Telescope in the 3.6, 4.5, 5.8 and 8.0 microns bands. The 
pre-reduced images are accessible in the Spitzer archive, and the final 
reduction steps are summarized in \citet{richard11}. Similarly to what we 
describe above, we convolved the NICMOS2 images with the PSF of the IRAC 
frames, and measured the mean $\rm NIC2/F160W - IRAC$ colors over the lensed 
images $\rm A2 + A3$, unresolved in the Spitzer images, using larger 
$3.0\arcsec$ diameter apertures. Because of the size and distance of the 
lensing galaxy (located at less than $4\arcsec$ from the midpoint between A2 
and A3), its contamination appeared in the form of an additional background 
level, nearly constant at the locations where the IRAC colors of $\rm A2 + A3$ 
were measured. This was removed when we made the photometric measurements. The 
comparison of the IRAC-PSF-convolved NIC2/F160W photometry with or without the 
central galaxy modeled and subtracted, however, leads to a contamination of 
less than 5\% in the overall photometry. The derived IR colors are listed in 
Table~\ref{tab:photometry}.

%

\subsection{Gravitational lens modeling}
\label{sect:reconstruction}

%

In order to derive accurate magnification factors and reconstruct the 
morphology of the 8~o'clock arc in the source plane, we had to correct for
distortions produced by the galaxy lens. To do so, we constructed the 
gravitational lens model of the system, using the public software
\texttt{LENSTOOL} \citep{kneib93,jullo07}. The high resolution of WFPC2 and 
NICMOS2 HST images allowed us to pin down precisely the bright centroid in each 
of the four detected counterpart images A1 to A4 of the lensed LBG. We used 
these multiple images as independent constraints on the mass distribution, 
assuming it follows a pseudo-isothermal elliptical profile (see 
\citet{limousin07} and \citet{richard10} for a precise description of this 
profile). We fixed the center of the mass distribution on the bright central 
galaxy of the system, but kept the ellipticity, $e$, position angle, $\theta$, 
velocity dispersion, $\sigma$, and core radius, $r_{\rm c}$, as free 
parameters. The best-fit parameters of the mass distribution are obtained with 
$e = 0.44\pm 0.16$, $\theta = 14\degr \pm 2\degr$ (east from the north), 
$\sigma = 348\pm 77\,\rm km\,s^{-1}$, and $r_{\rm c} = 
2.1^{+3.5}_{-0.2}\,\rm kpc$; the associated error bars are estimated from the 
range of models sampled by the Markov Chain Monte Carlo (MCMC) sampler 
\citep{jullo07}. The external critical line at the redshift of the 8~o'clock 
arc is overplotted in Fig.~\ref{fig:reconstruction} (left-hand panel). The best 
lens model yields an integrated mass within the Einstein radius 
($<3.32\arcsec$) of $1.96\times 10^{12}\,\rm M_{\sun}$, under the assumed 
cosmology. This is very close to the earlier lens modeling of the 8~o'clock arc 
made by \citet{allam07}, who found an integrated mass of 
$1.93\times 10^{12}\,\rm M_{\sun}$ within the same Einstein radius. Similarly, 
the authors found a comparable central velocity dispersion $\sigma = 
390\pm 10\,\rm km\,s^{-1}$, but with an error certainly underestimated because 
of the specific model (single isothermal elliptical) they assumed.

The best-fit model allowed us to estimate the overall magnification factors, 
$\mu$, for each image A2 and A3 targeted with the X-shooter spectrograph. These 
values are given in Table~\ref{tab:properties}. More importantly, the best-fit 
model could then be used to derive the geometrical transformation necessary for 
mapping the image plane coordinates into the source plane, and hence 
reconstruct the morphology of the 8~o'clock arc seen in the HST images in the 
source plane at $z=2.7350$. A reconstructed WFPC2/F450W-F650W-F814W 
composite-color image is shown in Fig.~\ref{fig:reconstruction} (right-hand 
panel). We clearly see that the source is formed of two major parts, the main 
galaxy component and a smaller blob separated by 0.15\,arcsec, or equivalently 
1.2\,kpc in projected distance. The corresponding lensed images of both the 
main component and the blob are indicated in Fig.~\ref{fig:reconstruction} 
(left-hand panel). It is certainly the higher magnification factor of the blob 
(twice the one of the main galaxy component) that enables us to resolve it in 
the source plane. The respective sizes (half-light radii) of the main galaxy 
component and the blob are $1.8\pm 0.2\,\rm kpc$ and $0.53\pm 0.05\,\rm kpc$.
Their errors were estimated by running SExtractor in a sample of source plane 
realizations, sampling the MCMC parameters of the lens model.

%

\begin{table*}
\caption{Nebular emission lines identified in the 8~o'clock arc images A2 and 
A3 with their fluxes}             
\label{tab:fluxes}
\centering                
\begin{tabular}{l c | c c c | c | l}
\hline\hline    
     &                                             & \multicolumn{3}{c|}{A2} & \multicolumn{1}{c|}{A3} & \\
Line & $\lambda_{\rm lab}$\,(\AA)\tablefootmark{a} & \multicolumn{1}{c}{$F_{\rm tot}$} & \multicolumn{1}{c}{$F_{\rm main}$} & \multicolumn{1}{c|}{$F_{\rm blob}$} & \multicolumn{1}{c|}{$F_{\rm tot}$} & Comments \\
\hline              
$[$\ion{O}{ii}]	                          & 3727.0897     &  $32.9\pm 2.7$ &  $20.7\pm 2.0$  &  $12.3\pm 1.5$ &  $24.9\pm 5.2$  & Strongly affected by telluric absorption \\
$[$\ion{O}{ii}]	                          & 3729.8804     &  $29.1\pm 1.8$ &  $18.3\pm 1.5$  &  $10.8\pm 1.1$ &  $22.4\pm 2.2$  & Strongly affected by telluric absorption \\
$[$\ion{Ne}{iii}]	                  & 3869.8468     &  $11.0\pm 0.6$ &   $6.9\pm 0.5$  &   $4.1\pm 0.4$ &   $6.9\pm 0.7$  & Partly affected by telluric absorption \\
H$\delta$	                          & 4102.8976     &   $8.5\pm 0.6$ &   $5.4\pm 0.5$  &   $3.3\pm 0.7$ &   $5.0\pm 0.7$  & Partly affected by sky residuals \\
H$\gamma$	                          & 4341.6903     &  $17.9\pm 0.4$ &  $11.3\pm 1.5$  &   $6.7\pm 1.4$ &  $12.7\pm 0.6$  & \\
H$\beta$	                          & 4862.6880     &  $46.0\pm 2.7$ &  $28.9\pm 2.1$  &  $17.1\pm 1.9$ &  $27.2\pm 2.7$  & Strongly affected by telluric absorption \\
H$\beta_{\rm inferred}$\tablefootmark{b}  &               &  $44.5\pm 1.8$ &  $28.4\pm 5.0$  &  $16.1\pm 4.4$ &  $31.3\pm 2.8$  & \\
$[$\ion{O}{iii}]	                  & 4960.2939     &  $50.5\pm 4.7$ &  $31.7\pm 3.1$  &  $18.8\pm 2.7$ &  $27.3\pm 5.7$  & Strongly affected by telluric absorption \\
H$\alpha$	                          & 6564.6329     & $181.2\pm 9.6$ & $120.7\pm 11.2$ &  $60.5\pm 8.4$ & $125.8\pm 17.7$ & Partly affected by telluric absorption \\
H$\alpha_{\rm intrinsic}$\tablefootmark{c}&               & $452.2\pm 72.2$& $336.2\pm 130.0$& $123.5\pm 73.2$& $302.5\pm 125.3$& \\
$[$\ion{N}{ii}]	                          & 6585.2284     &  $19.5\pm 3.7$ &  $15.6\pm 3.8$  &   $4.0\pm 2.7$ &  $20.1\pm 5.0$  & Noisy ($2-3\,\sigma$ detection) \\
\hline                                
\end{tabular}
\tablefoot{
Integrated line fluxes in units of $10^{-17}\,\rm erg\,s^{-1}\,cm^{-2}$. \\
For A2 are listed the total fluxes, $F_{\rm tot}$, and the individual fluxes 
corresponding to the decomposition of the line profiles into two Gaussian 
components associated with the main galaxy, $F_{\rm main}$, and a star-forming 
blob, $F_{\rm blob}$, as described in Sect.~\ref{sect:nebularlines}. \\
\tablefoottext{a}{Vacuum (laboratory) rest-frame wavelengths.} \\
\tablefoottext{b}{Inferred H$\beta$ line fluxes from the H$\gamma$ line flux 
and the $E(B-V)_{\rm gas}$ color excess derived from the H$\alpha$/H$\gamma$ 
Balmer decrement (Sect.~\ref{sect:dustextinction}).} \\
\tablefoottext{c}{Intrinsic H$\alpha$ line fluxes after correcting for dust 
extinction by $E(B-V)_{\rm gas}$ values derived from the H$\alpha$/H$\gamma$ 
Balmer decrement (Sect.~\ref{sect:dustextinction}).}
}
\end{table*}
%

\begin{table*}
\caption{Parameters of the Gaussian two-component fits for the 8~o'clock arc 
images A2 and A3 as constrained from the H$\gamma$ profile}
\label{tab:paramGaussians}
\centering                
\begin{tabular}{c | c c | c c}  
\hline\hline    
          & \multicolumn{2}{c|}{A2}                                    & \multicolumn{2}{c}{A3} \\
Component & $z_{\rm em}$ & $\sigma\,(\rm km\,s^{-1})$\tablefootmark{a} & $z_{\rm em}$ & $\sigma\,(\rm km\,s^{-1})$\tablefootmark{a} \\
\hline              
1 & $2.73423\pm 0.00012$ & $67\pm 7$ & $2.73357\pm 0.00040$ & $68\pm 30$ \\ 
2 & $2.73584\pm 0.00009$ & $45\pm 4$ & $2.73565\pm 0.00016$ & $61\pm 9$ \\
\hline                                
\end{tabular}
\tablefoot{ 
\\
\tablefoottext{a}{After subtracting in quadrature the instrumental resolution.}
}
\end{table*}
%

\begin{figure*}[!]
\centering
\includegraphics[width=7.5cm,clip]{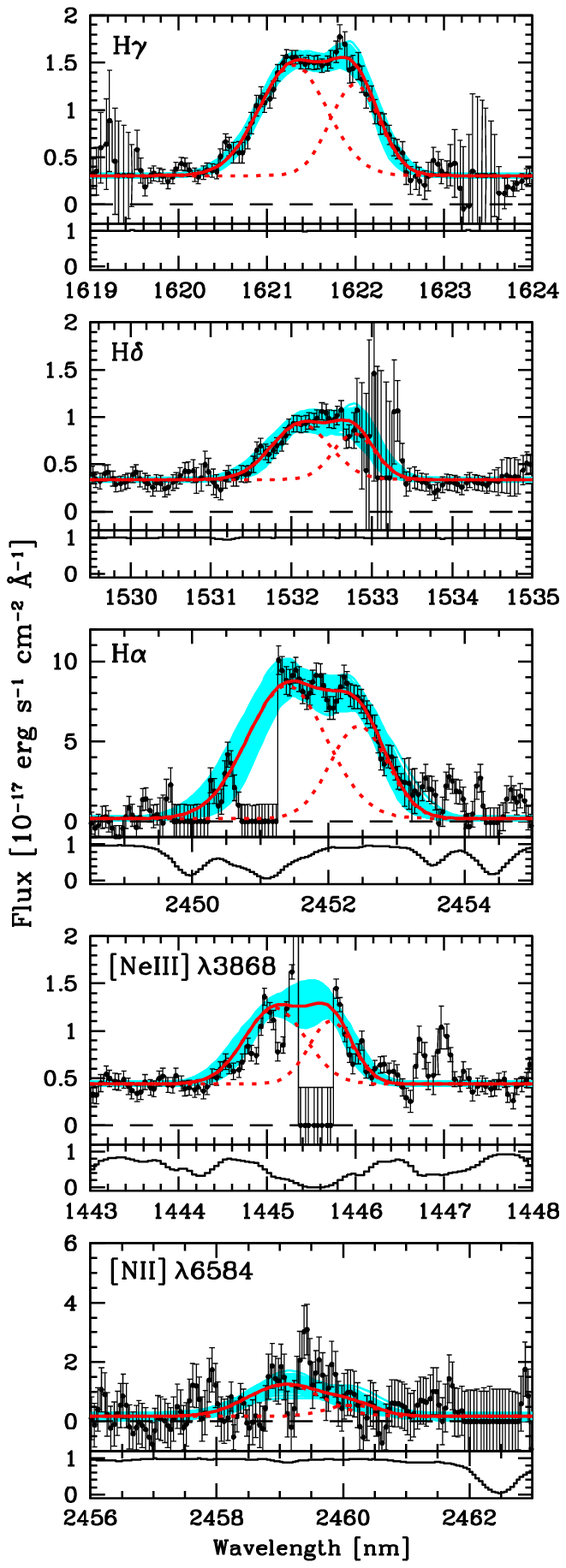}\hspace{0.5cm}
\includegraphics[width=7.5cm,clip]{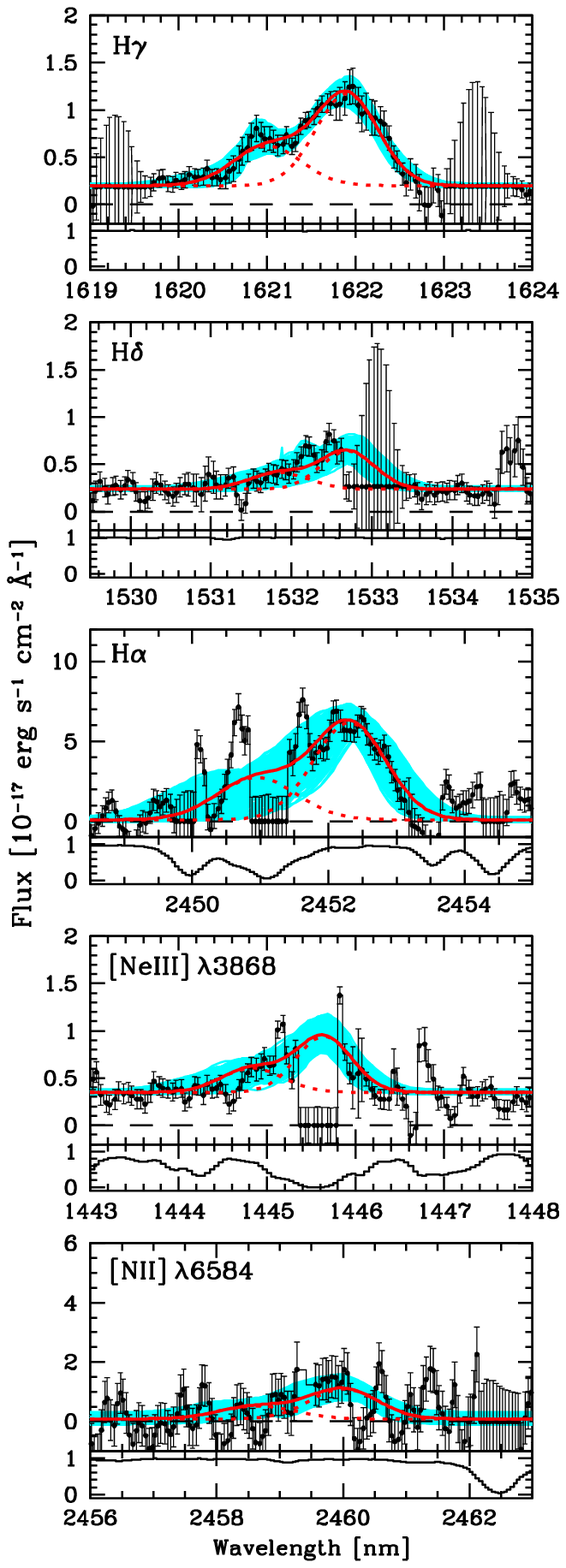}
\caption{Selection of nebular emission lines, free from strong atmospheric 
absorption, detected in the lensed image A2 ({\it left}-hand panels) and in the 
lensed image A3 ({\it right}-hand panels) of the 8~o'clock arc. In all panels, 
the black histogram represents the flux-calibrated data with $1\,\sigma$ errors 
as a function of vacuum-heliocentric-corrected wavelengths. The red continuous 
line is the best-fitted profile, and the red dotted line shows the fit of the 
two best individual Gaussian components with the respective parameters given in 
Table~\ref{tab:paramGaussians}. The cyan shaded area represents 68\% of the 
Monte Carlo runs, generated from the perturbation of the observed spectrum with 
a random realization of the error spectrum. The respective atmospheric 
transmission is plotted below each panel. The flux is set to zero (with large 
error bars) whenever the atmospheric transmission falls below 10\%.}
\label{fig:neblinesA2A3}
\end{figure*}
%

\begin{figure}[!]
\centering
\includegraphics[width=9cm,clip]{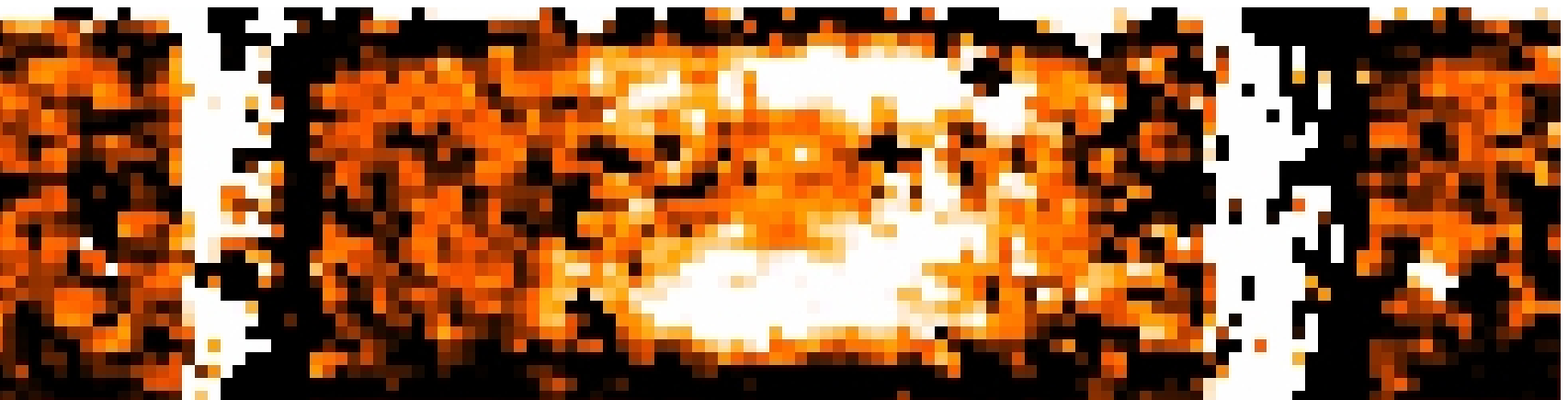}
\caption{Zoom on the wavelength-calibrated rectified sky-subtracted 2D spectrum 
around the nebular emission line H$\gamma$ plotted between 1618 and 1625\,nm. 
The wavelength scale grows from left to right, and the pixel scale 
perpendicular to the dispersion direction is equal to $0.2\arcsec$/pixel. 
Residuals of two sky lines are observed on both sides of the H$\gamma$ 
emission. The lower trace corresponds to the H$\gamma$ emission from A2 and the 
upper trace to the H$\gamma$ emission from A3. In the lensed image A2, two 
emission components are clearly distinguishable in the 2D H$\gamma$ profile, 
separated in the spectral direction by $\sim 130\,\rm km\,s^{-1}$ and in the 
spatial direction by $\sim 1\arcsec$. In the lensed image A3, one can also 
perceive two emission components, separated in the spectral direction only and 
with the bluest one being significantly weaker.}
\label{fig:2Dspectrum}
\end{figure}
%

\subsection{Nebular emission lines}
\label{sect:nebularlines}

%

Our X-shooter spectra cover two of the four lensed images of the 8~o'clock arc, 
a star-forming galaxy at the systemic redshift $z_{\rm sys} = 2.7350\pm 0.0003$ 
(DZ10). They exhibit a rich rest-frame UV spectrum studied in detail by DZ10, 
and a rich rest-frame optical spectrum with a number of nebular emission lines 
superposed on a weak continuum, which we describe below. The achieved S/N in 
the NIR allows us to analyze the spectra of the images A2 and A3 independently. 
Emission lines blueward of H$\gamma$ were previously not reported by 
\citet{finkelstein09}, because they are not covered by their spectra. In 
Table~\ref{tab:fluxes} we list the detected nebular emission lines in A2 and 
A3. A selection of them, free from atmospheric absorption lines, is reproduced 
in Fig.~\ref{fig:neblinesA2A3}. 

As seen in Fig.~\ref{fig:neblinesA2A3}, the nebular lines of the 8~o'clock arc 
show asymmetric profiles, suggesting they consist of more than one component, 
and more surprisingly, they show different profiles between the lensed images 
A2 and A3. Inspection of the 2D spectra (see Fig.~\ref{fig:2Dspectrum} showing 
the 2D H$\gamma$ profile) confirms the presence of mainly two components and a 
difference in the emission distribution between A2 and A3. In the image A2 the 
two components are clearly distinguishable, being separated by 
$\sim 130\,\rm km\,s^{-1}$ in the spectral direction and spatially shifted by 
$\sim 1\arcsec$, while in the image A3 we observe two components with a very 
similar separation in the spectral direction, but with the bluer component that
is significantly weaker and without spatial separation. We are tempted to 
interpret the redder, spatially shifted component of A2 as the spectral 
signature of the star-forming blob identified in the reconstructed source plane 
(Sect.~\ref{sect:reconstruction}). Indeed, with a magnification factor more 
than twice higher than that of the main galaxy, we may expect that its emission 
is sufficiently boosted by gravitational lensing to be spectroscopically 
detected. The blob should also have a counterpart image toward A3 with a 
similar magnification factor $\mu = 11.9$ (Fig.~\ref{fig:reconstruction}, 
left-hand panel). The redder component of A3 might correspond to the emission 
from the blob, because it is almost at an identical velocity to that of the 
redder component in A2 (see Table~\ref{tab:paramGaussians}). To explain the 
lack of its spatial shift and its significantly larger strength relative to the 
bluer component, while the redder and bluer components are roughly the same 
strength in A2, we can possibly blame the slit orientation, which did not 
optimally cover the lensed image A3, resulting in a partial loss of the 
spectral information, either on the blob's or main galaxy's spatial and 
velocity structure. As an alternative explanation, we can advocate a local 
magnification of the image A2 owing to the presence of substructure in the lens 
plane, which makes it looks a bit more extended/brighter compared to the other 
lensed image A3. Follow-up observations are necessary to distinguish between 
these two possibilities and to understand the complex velocity structure of 
images A2 and A3 in the rest-frame optical spectra.

The nebular emission lines were analyzed with a multi-Gaussian fitting 
procedure based on the non-linear $\chi^2$ minimization and the 
Levenberg-Marquardt algorithm. Windows of the NIR spectrum with atmospheric 
transmissions falling below 10\% were excluded from the multi-Gaussian fitting. 
Varying the number of possible Gaussian components in the fits, a combination 
of two Gaussian profiles yielded the best-fit solution for all observed line 
profiles. H$\gamma$, the most reliable line, which is free from telluric 
absorption and sky residuals, was used to constrain the best-fit values of 
redshifts (Gaussian centroids), $z_{\rm em}$, and velocity 
dispersions\footnote{The velocity dispersion is related to the width of a line 
through $\sigma = \rm FWHM/2.355\times c/\lambda_{\rm obs}$, where FWHM is the 
full-width-at-half maximum in wavelength of a line at $\lambda_{\rm obs}$.} 
(Gaussian widths), $\sigma$, of the two Gaussian functions. We also used it for 
the fitting of all other nebular lines; only amplitudes were allowed to vary. 
H$\beta$, [\ion{O}{iii}]\,$\lambda$4960 and the [\ion{O}{ii}] doublet, all 
heavily affected by atmospheric absorption lines, were even fitted by fixing 
the amplitude ratio of the two Gaussian profiles to that of the H$\gamma$ 
profile. Errors on the values of $z_{\rm em}$, $\sigma$ and fluxes were 
estimated using a Monte Carlo approach, whereby the observed spectrum was 
perturbed with a random realization of the error spectrum and refitted. The 
process was repeated 1000 times and the error in each quantity was taken to be 
the standard deviation of the values generated by the 1000 Monte Carlo runs. 

We obtained a satisfactory fit for all nebular lines, which demonstrates the 
robustness of the procedure and the reliability of the H$\gamma$ profile as  
calibrator. The corresponding best-fitted Gaussian profiles with their errors 
are shown in Fig.~\ref{fig:neblinesA2A3} with red continuous lines and 
cyan-shaded areas that delimit 68\% of the generated Monte Carlo runs. 
Moreover, the best fit obtained in this way for the blend of the light that 
comes from the main galaxy and the blob in A2 agrees within $1\,\sigma$ with 
the results we obtained when we separately and manually extracted the two 
corresponding signals from the 2D H$\gamma$ profile. Given the limited S/N of 
our observations, we believe that a single extraction from the 2D spectrum and 
a multi-Gaussian treatment of the blended components leads to the most reliable 
results.

In Table~\ref{tab:paramGaussians} we list the best values of redshifts and 
velocity dispersions as constrained from the H$\gamma$ profile for the two 
Gaussian functions that were used to fit all observed nebular lines of the 
8~o'clock arc images A2 and A3. Averaging the four redshifts of the two 
Gaussian functions of A2 and A3 yields a mean redshift of the ionized gas of 
$\langle z_{\rm em}\rangle = 2.7348\pm 0.0005$, which excellently agrees with 
$z_{\rm sys} = 2.7350\pm 0.0003$ determined for the 8~o'clock arc by DZ10 from 
photospheric absorption lines and emission lines detected in the rest-frame UV 
spectrum. Our wavelength calibration solution was, in addition, cross-checked 
over the entire NIR spectral range by fitting Gaussian functions to the more 
than 400 sky lines listed in \citet{rousselot00}. We obtained a median accuracy 
of the wavelength solution of 0.2\,\AA, i.e., half a pixel, or 
$\sim 10\,\rm km\,s^{-1}$. This definitely confirms an offset of about 
$225\,\rm km\,s^{-1}$ between the systemic redshift as determined from 
X-shooter spectra and by \citet[][$z_{\rm sys} = 2.7322\pm 0.0012$]
{finkelstein09}. The superiority of our data because of a 10~$\times$ higher 
spectral resolution clearly makes our redshift determination more reliable.

The measured nebular line fluxes are listed in Table~\ref{tab:fluxes}. We 
provide the respective individual fluxes of the two Gaussian profiles plus the 
total fluxes (sum of the two profiles) for the image A2 and the total fluxes
only for the image A3. Our total line fluxes derived for H$\alpha$ and 
H$\gamma$ agree very well with the measurements obtained by 
\citet{finkelstein09} for both the images A2 and A3. For H$\beta$, 
[\ion{O}{iii}]\,$\lambda$4960, as well as [\ion{N}{ii}]\,$\lambda$6585, the 
discrepancy is large, on the other hand. The H$\beta$ and [\ion{O}{iii}] lines 
are heavily affected by atmospheric absorption lines in the $J$ and $H$ bands 
and the [\ion{N}{ii}] line is relatively weak, which makes their respective 
flux measurement particularly difficult. Benefiting from higher quality NIR 
spectra than those of Finkelstein et~al., our flux measurements are expected to 
be more accurate.

%

\begin{table*}
\caption{Physical properties derived for the 8~o'clock arc images A2 and A3}    
\label{tab:properties}
\centering                
\begin{tabular}{l | c c c | c | l}  
\hline\hline    
 & \multicolumn{3}{c|}{A2} & \multicolumn{1}{c|}{A3} & \\
 & Total & Main & Blob     & Total                   & Reference \\
\hline
Magnification factor $\mu$                                & $5.0\pm 1.0$   & $5.0\pm 1.0$   & $11.7\pm 2.3$  & $3.9\pm 0.8$   & Sect.~\ref{sect:reconstruction} \\
Half-light radius $r_{1/2}\,(\rm kpc)$                    & $1.8\pm 0.2$   & $1.8\pm 0.2$   & $0.53\pm 0.05$ & $1.8\pm 0.2$   & Sect.~\ref{sect:reconstruction} \\
$E(B-V)_{\rm gas}$\tablefootmark{a}                       & $0.30\pm 0.04$ & $0.34\pm 0.10$ & $0.23\pm 0.15$ & $0.29\pm 0.09$ & Sect.~\ref{sect:dustextinction} \\
$E(B-V)_{\beta}$\tablefootmark{b}                         & $0.23\pm 0.10$ &                &                & $0.16\pm 0.10$ & Sect.~\ref{sect:dustextinction} \\
$E(B-V)_{\rm stars}$\tablefootmark{c,d}                   & & & \multicolumn{2}{c|}{$0.19\pm 0.04$}                           & Sect.~\ref{sect:SED} \\
$12+\log(\rm O/H)_{N2}$\tablefootmark{e}                  & $8.35\pm 0.19$ & $8.39\pm 0.19$ & $8.22\pm 0.25$ & $8.46\pm 0.19$ & Sect.~\ref{sect:metallicity} \\
$Z_{\rm N2}/Z_{\sun}$\tablefootmark{f}                    & $0.46\pm 0.20$ & $0.50\pm 0.22$ & $0.34\pm 0.20$ & $0.59\pm 0.26$ & Sect.~\ref{sect:metallicity} \\
$12+\log(\rm O/H)_{Ne3O2}$\tablefootmark{g}               & $>8.09$        & $>8.09$        & $>8.09$        & $>8.18$        & Sect.~\ref{sect:metallicity} \\
$\rm SFR_{H\alpha}\,(M_{\sun}\,yr^{-1})$\tablefootmark{h} & $279\pm 45$    & $207\pm 80$    & $33\pm 19$     & $239\pm 99$    & Sect.~\ref{sect:SFR} \\
$\rm SFR_{UV}\,(M_{\sun}\,yr^{-1})$\tablefootmark{i}      & $156\pm 110$   &                &                & $198\pm 140$   & Sect.~\ref{sect:SFR} \\
$\rm SFR_{SED}\,(M_{\sun}\,yr^{-1})$\tablefootmark{d}     & & & \multicolumn{2}{c|}{$162^{+124}_{-95}$}                       & Sect.~\ref{sect:SED} \\
$W_0(\rm H\alpha)\,(\AA)$\tablefootmark{j}                & $134\pm 67$    & $89\pm 45$     & $45\pm 23$     & $139\pm 72$    & Sect.~\ref{sect:EW} \\
$W_0(\rm H\alpha)\,(\AA)$\tablefootmark{k}                & $187\pm 124$   & $139\pm 115$   & $51\pm 49$     & $187\pm 152$   & Sect.~\ref{sect:EW} \\
$W_0(\rm H\beta)\,(\AA)$\tablefootmark{j}                 & $30\pm 2$      & $19\pm 3$      & $11\pm 3$      & $29\pm 3$      & Sect.~\ref{sect:EW} \\
$W_0(\rm H\beta)\,(\AA)$\tablefootmark{k}                 & $47\pm 14$     & $35\pm 22$     & $13\pm 12$     & $44\pm 23$     & Sect.~\ref{sect:EW} \\
$W_0(\rm H\alpha)_{SED}\,(\AA)$\tablefootmark{d}          & & & \multicolumn{2}{c|}{$255\pm 95$}                              & Sect.~\ref{sect:SED} \\
$\rm Age\,(Myr)$\tablefootmark{d}                         & & & \multicolumn{2}{c|}{$40^{+25}_{-20}$}                         & Sect.~\ref{sect:SED} \\
$M_{\rm stars}\,(\rm 10^9\,M_{\sun})$\tablefootmark{d}    & & & \multicolumn{2}{c|}{$7.9^{+2.5}_{-2.0}$}                      & Sect.~\ref{sect:SED} \\ 
$M_{\rm gas}\,(\rm 10^9\,M_{\sun})$\tablefootmark{l}      & $20.9\pm 3.9$  & $17.6\pm 5.6$  & $2.2\pm 0.9$   & $19.1\pm 6.4$  & Sect.~\ref{sect:gasmass} \\
$M_{\rm dyn,rot}\,(\rm 10^9\,M_{\sun})$\tablefootmark{m}  & $16.0\pm 4.9$  & $8.4\pm 2.7$   & $1.1\pm 0.3$   & $20.2\pm 15.9$ & Sect.~\ref{sect:dynmass} \\
$M_{\rm dyn,disp}\,(\rm 10^9\,M_{\sun})$\tablefootmark{n} & $14.3\pm 4.5$  & $12.6\pm 4.0$  & $1.7\pm 0.5$   &                & Sect.~\ref{sect:dynmass} \\
\hline
\end{tabular}
\tablefoot{
For A2 are listed the total values and the individual values corresponding to 
the main galaxy and the star-forming blob, as described in 
Sects.~\ref{sect:reconstruction} and \ref{sect:nebularlines}. \\
\tablefoottext{a}{Gas-phase (nebular) dust extinctions computed from the 
H$\alpha$/H$\gamma$ Balmer decrement.} \\ 
\tablefoottext{b}{Stellar dust extinctions derived from the UV slopes, $\beta$, 
measured from the observed $(V-I)$ colors (Table~\ref{tab:photometry}), 
following the prescriptions of \citet{bouwens09}.}\\
\tablefoottext{c}{Stellar dust extinction derived from the SED modeling and 
corrected for the Galactic dust extinction, $E(B-V)^{\rm Gal} = 0.056$, at the 
position of the 8~o'clock arc \citep{schlegel98}.} \\
\tablefoottext{d}{Combined SED modeling results obtained with and without the 
treatment of nebular emission (continuum plus lines) and scaled to the 
\citet{chabrier03} IMF. The predicted $W_0(\rm H\alpha)_{SED}$ comes, of 
course, only from SED models including the nebular emission.} \\
\tablefoottext{e}{Oxygen abundances derived from the N2 calibration from
\citet{pettini04}. The errors include the 0.18\,dex systematic uncertainty in 
the N2 calibration zeropoint.} \\
\tablefoottext{f}{Gas-phase metallicities relative to the solar value of 
$12+\log(\rm O/H)_{\sun} = 8.69$ \citep{asplund09}.} \\
\tablefoottext{g}{Oxygen abundances derived from the Ne3O2 calibration from
\citet{nagao06}.} \\
\tablefoottext{h}{Star-formation rates computed from the H$\alpha$ luminosity 
and corrected for dust extinction by $E(B-V)_{\rm gas}$ values and for 
gravitational lensing by the magnification factor $\mu$.} \\
\tablefoottext{i}{Star-formation rates computed from the UV continuum 
luminosity at the rest-frame wavelength of 1600\,\AA\ and corrected for dust 
extinction by $E(B-V)_{\rm stars}$ values and for gravitational lensing by the 
magnification factor $\mu$.} \\
\tablefoottext{j}{Rest-frame equivalent widths computed from the observed 
Balmer line and local continuum fluxes, assuming that the same dust extinction 
applies to the nebular and stellar emission.} \\
\tablefoottext{k}{Rest-frame equivalent widths computed from the dust-corrected 
Balmer line and local continuum fluxes, assuming that the gas-phase dust 
extinction, $E(B-V)_{\rm gas}$, applies to the nebular emission and the stellar 
dust extinction, $E(B-V)_{\rm stars}$, applies to the stellar continuum.} \\
\tablefoottext{l}{Gas masses derived from the H$\alpha$ star-formation rates, 
$\rm SFR_{H\alpha}$, through the Schmidt-Kennicutt relation. The rest-frame UV 
star-formation rates, $\rm SFR_{UV}$, corrected for dust extinction by 
$E(B-V)_{\rm stars}$ values, lead to smaller gas masses by a factor of up to
$\sim 1.4$.} \\
\tablefoottext{m}{Dynamical masses derived under the assumption of 
rotation-dominated kinematics.} \\
\tablefoottext{n}{Dynamical masses derived under the assumption of 
dispersion-dominated kinematics.}
}
\end{table*}
%

\section{Physical properties}
\label{sect:properties}

Thanks to the magnification provided by gravitational lensing, we detected a 
larger number of nebular emission lines in the 8~o'clock arc and this with a 
higher S/N than what is typically seen in unlensed high-redshift objects. These 
emission line fluxes which are collected in Table~\ref{tab:fluxes} and their 
ratios allow us to probe several physical properties that characterize the 
ionized gas of the lensed LBG and the galaxy itself. The multi-wavelength 
photometry (Table~\ref{tab:photometry}) also provides complementary valuable 
physical quantities on the galaxy via the spectral energy distribution (SED) 
modeling. In the following, we make the assumption that the gas in the 
8~o'clock arc is ionized by OB stars, with negligible contributions from an 
active galactic nucleus (AGN), given the low [\ion{N}{ii}]/H$\alpha$ line 
ratio of $\sim 0.15$. Table~\ref{tab:properties} lists all the physical 
properties derived as described in the sections below. For the lensed image A2, 
we separately report the physical quantities corresponding to the total A2 line 
profile (column~2) and to the decomposition of the A2 line profile into the 
main galaxy component (column~3) and the smaller blob (column~4), as identified 
in the reconstructed source plane (Sect.~\ref{sect:reconstruction}). For the 
lensed image A3, we report only the physical quantities corresponding to the 
total A3 line profile (column~5).

%

\begin{figure}[!]
\centering
\includegraphics[width=9cm,clip]{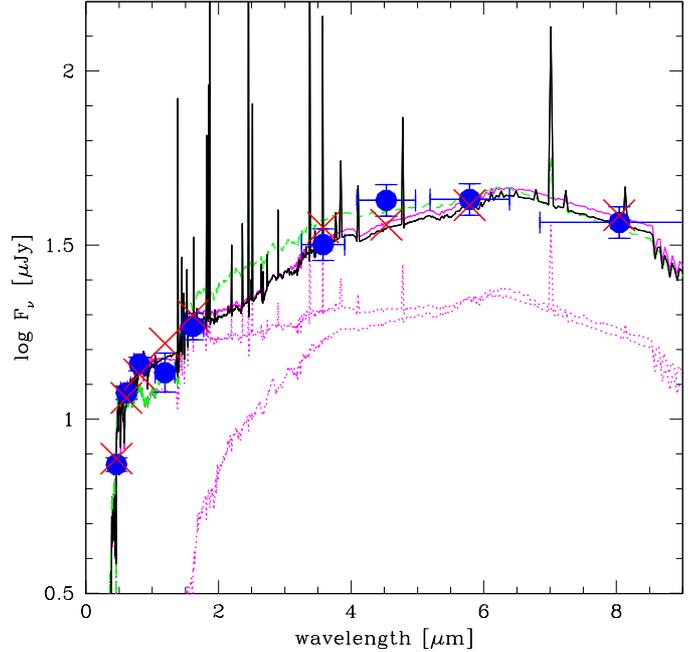}
\caption{Observed SED of the 8~o'clock arc with the photometry of the lensed 
image A2 not corrected for gravitational lensing (blue points with error bars). 
Best-fit SED models obtained with templates with nebular emission (black line 
and red crosses showing the synthetic fluxes in the broad-band filters) and 
without nebular emission (magenta line). The dotted magenta lines show the 
contributions of a two-population fit, assuming a maximally old stellar 
population ($\sim 2.3\,\rm Gyr$ for an instantaneous burst), plus a younger
stellar population with a variable age and star-formation history. The sum of 
the two (green dashed line) also provides a reasonable fit. Its stellar mass,
however, significantly exceeds the dynamical mass.}
\label{fig:SED}
\end{figure}
%

\subsection{SED modeling}
\label{sect:SED}

%

To estimate the physical properties of the stellar population of the 8~o'clock 
arc, we fitted the observed broad-band SED with our fitting tool described in 
\citet{schaerer09,schaerer10}. To estimate uncertainties on the fit parameters, 
we used Monte Carlo simulations with typically 1000 realizations of the data. 
The images A2 and A3 were fitted separately, using the photometry given in 
Table~\ref{tab:photometry} after correcting for magnification by factors
$\mu^{\rm A2} = 5.0$ and $\mu^{A3} = 3.9$, respectively. Because A2 and A3 are 
different images of the same galaxy and because their (lensing-corrected) 
photometry is quite consistent, we determined the physical properties of this 
object from the total probability distribution function of the two lensed 
images. The indicated uncertainties are derived from the 68\% confidence 
interval.

In practice, the SED fits are carried out with the following assumptions and 
templates, and varying the following parameters. We adopted the 
\citet{bruzual03} spectral templates with solar metallicity, assuming a 
\citet{salpeter55} initial mass function (IMF) from 0.1 to $100\,\rm M_{\sun}$. 
We allowed for 10 exponentially declining star-formation histories with 
e-folding times between $30\,\rm Myr$ and $3\,\rm Gyr$ and for a constant 
star-formation rate. Nebular emission (lines and continuum) was optionally 
included, following the \citet{schaerer09,schaerer10} prescriptions. The 
Ly$\alpha$ emission line flux was set to zero, as observed and understood from 
radiation transfer models (see DZ10). Extinction, described by the 
\citet{calzetti00} attenuation law, was varied from $A(V) = 0$ to $2\,\rm mag$, 
in steps of $0.05\,\rm mag$. A Galactic foreground extinction $E(B-V)^{\rm Gal} 
= 0.056$ was taken into account \citep{schlegel98}. The redshift is fixed by 
the spectroscopic observations.

The derived fitting parameters are the visual extinction, $A(V)$, the age of 
the stellar population (defined as the age since the onset of star formation), 
the stellar mass, $M_{\rm stars}$, and the current star-formation rate, 
$\rm SFR_{\rm SED}$. When nebular emission is included, we also predict 
the H$\alpha$ rest-frame equivalent width, $W_0(\rm H\alpha)_{SED}$, which can 
be compared to values obtained from our observations. The 1D confidence level 
for each physical parameter is derived by marginalization over all other 
parameters.

Good fits are obtained with and without nebular emission, as shown in 
Fig.~\ref{fig:SED}, and the respective overall physical properties are quite 
similar. The resulting ages and star-formation timescales are comparable, 
implying that the derived parameters are similar to values obtained if we 
assume a constant SFR. Typically, we find, within 68\% confidence level, 
$A(V) \sim 0.55-1\,\rm mag$, ages of $\sim 20-65\,\rm Myr$, stellar masses of 
$M_{\rm stars} \sim (6-11)\times 10^9\,\rm M_{\sun}$, and star-formation rates 
of $\rm SFR_{SED} \sim 55-340\,M_{\sun}\,yr^{-1}$ from models with and without 
nebular emission\footnote{The stellar masses and star-formation rates were 
corrected by the factor of 1.8 to account for the flattening of the 
\citet{chabrier03} IMF compared to the \citet{salpeter55} IMF, and have 
consistent results with those of Sects.~\ref{sect:SFR} and \ref{sect:masses}.}. 
The main differences between SED models with and without nebular emission are 
observed in ages and dust extinctions; they remain compatible within 68\% 
confidence level, however. For example, we find $A(V) = 
0.87^{+0.13}_{-0.12}\,\rm mag$ for standard templates and $A(V)_{\rm stars} = 
0.65\pm 0.10\,\rm mag$ with nebular emission. The latter fits may indicate a 
somewhat lower extinction than derived from the Balmer decrement (see 
Sect.~\ref{sect:dustextinction}), but the difference is not significant, as
discussed in Sect.~\ref{sect:dustcorrections}. The predicted H$\alpha$ 
rest-frame equivalent width of the model with nebular emission, 
$W_0(\rm H\alpha) = 255\pm 95$\,\AA, is slightly too large compared to the 
observations (see Sect.~\ref{sect:EW}).

Comparing our results to the SED fits obtained by \citet{finkelstein09}, we 
note that our stellar mass estimate is significantly lower than theirs, 
$M_{\rm stars} \sim 4.2\times 10^{11}\,\rm M_{\sun}$ determined with a 
Salpeter IMF, by more than one order of magnitude. Their photometry comes from 
the SDSS, plus $H$ and $K^\prime$ images from the Near InfraRed Imager and 
Spectrometer (NIRI) on the Gemini North telescope, whereas we used the more 
reliable HST and Spitzer photometry up to 8.0 microns. Assuming that the SEDs 
shown in their Fig.~7 correspond to observed, lensing-corrected fluxes, it 
appears that the NIR fluxes (for instance in the $H$ band) agree quite well for 
the image A2, but for the image A3 their $H$-band flux is nearly twice as high 
as ours (assuming a magnification factor $\mu^{\rm A3} = 3.9$). Moreover, they 
incorrectly determine the galaxy stellar mass by summing the contributions from 
images A2 and A3, whereas A2 and A3 represent two lensed images of the same 
object. The main difference probably comes from their preference for a 
two-burst stellar population model with a maximally old component, which yields 
stellar masses of $M_{\rm stars} \sim (1-3)\times 10^{11}\,\rm M_{\sun}$ for 
images A2 and A3. We have investigated to which extent an old stellar 
population ($\sim 2.3\,\rm Gyr$ for an instantaneous burst) may be present. The 
IRAC fluxes from 3.6 to $8.0\,\mu \rm m$ are, indeed, compatible with such a 
population, which accounts for up to $\sim 50$\% of the IRAC fluxes, as shown 
in Fig.~\ref{fig:SED}. In this case, we obtain a stellar mass as large as 
$M_{\rm stars} \sim 1.8\times 10^{11}\,\rm M_{\sun}$ for the old population,
similarly to \citet{finkelstein09}, plus a $7-8\times$ lower mass for the 
younger population. There is not much more room for a larger contribution of 
the old population because otherwise the NIR SED ($\la 2\,\mu \rm m$) is 
underpredicted by the young population. Nevertheless, such a large stellar mass 
is not compatible with our dynamical mass estimate (see 
Sect.~\ref{sect:dynmass}), because it should not exceed the dynamical mass. In 
summary, we exclude the presence of a dominating old stellar population. With 
a stellar mass of $M_{\rm stars} \sim (1-2)\times 10^{10}\,\rm M_{\sun}$ (with 
the Salpeter IMF), derived from our SED fits based on recent WFPC2, NICMOS2 and 
IRAC observations onboard HST and Spitzer and using our well-tested fitting 
tools, the 8~o'clock galaxy falls close to the median of the $M_{\rm stars}$ 
distribution of $z\sim 2$ star-forming galaxies \citep{erb06b}.

%
%


%

\subsection{Dust extinction}
\label{sect:dustextinction}

%

The dust extinction is best derived from the Balmer decrement because the 
Balmer lines have well constrained flux ratios from statistical equilibrium 
calculations. Any deviation from the theoretical values are attributed to the
gas-phase (nebular) dust reddening. In the 8~o'clock arc, the 
H$\alpha$/H$\gamma$ ratio leads to the most reliable dust extinction estimate 
because these two lines are the least affected by telluric absorptions and sky 
residuals. As is common practice in the analysis of \ion{H}{ii} regions, we 
assumed Case B recombination, an electron temperature $T_{\rm e} = 
10^4\,\rm K$, and electron densities in the range $n_{\rm e} = 
10^2-10^4\,\rm cm^{-3}$ for the intrinsic Balmer line ratios 
\citep{osterbrock89}. The corresponding intrinsic H$\alpha$/H$\gamma$ ratio is 
equal to 6.159. Considering the \citet{calzetti00} starburst reddening curve, 
the observed H$\alpha$/H$\gamma$ ratios imply gas-phase color excesses 
$E(B-V)_{\rm gas}^{\rm A2} = 0.30\pm 0.04$ for the image A2 and 
$E(B-V)_{\rm gas}^{\rm A3} = 0.29\pm 0.09$ for the image A3. We stress that the 
correction of the Galactic extinction, $E(B-V)^{\rm Gal} = 0.056$, at the 
position of the LBG \citep{schlegel98}, is negligible at the observed NIR 
wavelengths of Balmer lines. 

The respective gas-phase color excesses of images A2 and A3 agree very well, as 
do their respective $V-I$ colors (see Table~\ref{tab:photometry}). Indeed, the 
UV slopes, $\beta$, measured from the observed $(V-I)$ colors, following the 
prescriptions of \citet{bouwens09}, are $\beta = -1.20\pm 0.43$ 
($-1.50\pm 0.43$) for A2 (A3). This translates to $E(B-V)_{\beta} = 
0.23\pm 0.10$ ($0.16\pm 0.10$) for A2 (A3). As a result, the respective 
gas-phase and stellar dust extinctions of the two images A2 and A3 are very 
similar. This is what is expected because in the context of the gravitational 
lens modeling, the lensed images A2 and A3 represent physically the same 
regions. Our derived $E(B-V)_{\rm gas}$ excellently agrees with the color 
excess of A2 determined by \citet{finkelstein09} from the same 
H$\alpha$/H$\gamma$ ratio. However, they find a significant dust extinction 
difference between images A2 and A3, both on the basis of the 
H$\alpha$/H$\gamma$ ratio and broadband photometry (see their Table~3 and 
Fig.~7), which we here confirm, based on better data (X-shooter spectra and HST 
images), should not be the case.

The Gaussian decomposition of the A2 line profile into a first component 
associated with the main galaxy and a second one with the star-forming blob 
yields $E(B-V)_{\rm gas}^{\rm main} = 0.34\pm 0.10$ and 
$E(B-V)_{\rm gas}^{\rm blob} = 0.23\pm 0.15$, respectively. These values show a 
trend toward a smaller color excess in the blob, however, the trend is only 
marginal because all measurements are within $1\,\sigma$ errors. 

One important application of the determined dust extinction is for inferring 
fluxes of Balmer lines, when those are only partially or not detected. This is 
particularly interesting in the case of the 8~o'clock arc for the H$\beta$ 
line, whose profile is only partially detected because the atmospheric 
transmission falls locally below 10\%. The H$\beta$ flux can then be determined 
from the observed H$\gamma$ flux and our estimate of $E(B-V)_{\rm gas}$. The 
H$\beta$ fluxes so derived are listed in Table~\ref{tab:fluxes}. They are 
within $1\,\sigma$ from the measured values, a nice agreement that provides 
another demonstration of the robustness of our Gaussian fitting procedure. 

%

\subsection{Metallicity}
\label{sect:metallicity}

%

Nebular emission lines are commonly used to estimate the metallicity in
extragalactic \ion{H}{ii} regions. There are several empirical metallicity
calibrations, leading to oxygen abundance estimates, from various emission line 
ratios. Because we detected [\ion{N}{ii}] and [\ion{Ne}{iii}] in the 8~o'clock 
arc, we use the N2 index from \citet{pettini04}:
\begin{equation} \label{eq:N2}
12+\log({\rm O/H}) = 8.90+0.57\times 
    \log [F([\ion{N}{ii}]\,\lambda6585)/F({\rm H}\alpha)]\,,
\end{equation} 
and the Ne3O2 index from \citet{nagao06}:
\begin{eqnarray} \label{eq:NeIIIOII}
\log
[F([\ion{Ne}{iii}]\,\lambda3869)/(F([\ion{O}{ii}]\,\lambda3727)+F([\ion{O}{ii}]\,\lambda3729))] \nonumber \\
    = -82.202 + 32.014x - 4.0706x^2 + 0.16784x^3\,,
\end{eqnarray} 
where $x = 12+\log(\rm O/H)$. The two indices have the advantage of having a 
single-valued dependence on the oxygen abundance and are relatively robust with 
respect to flux calibration and dust extinction (because one uses ratios of 
lines that are close in wavelength). Given the uncertain flux measurements of 
both the [\ion{O}{ii}] and [\ion{O}{iii}] doublets, the popular $R_{23}$ index 
\citep{pagel79} is hardly applicable for the 8~o'clock arc.

We find from the N2 index $12+\log(\rm O/H)_{N2}^{A2} = 8.35\pm 0.19$ for the 
image A2 and $12+\log(\rm O/H)_{N2}^{A3} = 8.46\pm 0.19$ for the image A3. The 
two measurements agree very well within $1\,\sigma$ errors, and also agree with 
the metallicity derived by \citet{finkelstein09}, $12+\log(\rm O/H)_{N2} = 
8.58\pm 0.18$. From our data, we determine an average gas-phase metallicity 
$Z_{\rm N2}^{\rm A2,A3} = 0.53\pm 0.23\,Z_{\sun}$ for the 8~o'clock arc, when 
adopting a solar value of $12+\log(\rm O/H)_{\sun} = 8.69$ \citep{asplund09}. 
The oxygen abundances deduced from the Ne3O2 index, 
$12+\log(\rm O/H)_{Ne3O2}^{A2} > 8.09$ ($Z_{\rm Ne3O2}^{\rm A2} > 
0.25\,Z_{\sun}$) for the image A2 and $12+\log(\rm O/H)_{Ne3O2}^{A3} > 8.18$ 
($Z_{\rm Ne3O2}^{\rm A3} > 0.31\,Z_{\sun}$) for the image A3, are consistent 
with the N2 index results. They are considered as lower limits, because of the 
likely underestimation of the flux of the [\ion{O}{ii}] lines, heavily affected 
by atmospheric absorptions.

In comparison with the stellar metallicity, $Z_{\rm stars} = 0.82\,Z_{\sun}$, 
and the metallicity of the interstellar medium, $Z_{\rm ISM} = 0.65\,Z_{\sun}$, 
of the 8~o'clock arc determined in DZ10, our gas-phase metallicity estimate 
agrees well, keeping in mind that (i)~the [\ion{N}{ii}]\,$\lambda$6585 
line we use is detected at only $2-3\,\sigma$, being located at a wavelength 
where the detector noise significantly increases, and (ii)~the \ion{H}{ii} 
region metallicity calibrations (\ref{eq:N2}) and (\ref{eq:NeIIIOII}) have a 
systematic $1\,\sigma$ uncertainty of $\pm 0.2\,\rm dex$. Adopting 
$Z_{\rm mean} = 0.67\pm 0.23\,Z_{\sun}$, the mean of all the metallicity 
indicators available for the 8~o'clock arc, as the metallicity of this 
high-redshift LBG, seems to be a reasonable estimate.

Considering separately the contributions from the main galaxy and the
star-forming blob in the lensed image A2, we derive from the N2 index 
$Z_{\rm N2}^{\rm main} = 0.50\pm 0.22\,Z_{\sun}$ for the main component, while 
the blob shows a trend toward a lower metallicity with $Z_{\rm N2}^{\rm blob} 
= 0.34\pm 0.20\,Z_{\sun}$. This remains a speculative result, given the weak 
S/N over the [\ion{N}{ii}]\,$\lambda$6585 line profile.

%

\subsection{Star-formation rate}
\label{sect:SFR}

%

We can obtain an estimate of the star-formation rate (SFR) from the luminosity
in the H$\alpha$ emission line, through the calibration by \citet{kennicutt98}:
\begin{equation} \label{eq:SFR}
{\rm SFR_{H\alpha}\,(M_{\sun}\,yr^{-1})} = 
    7.9\times 10^{-42} L({\rm H}\alpha) 
    \times \frac{1}{1.8}\times \frac{1}{\mu}\times 1.1\,.
\end{equation}
The three correction factors (last three terms in (\ref{eq:SFR})) added to the 
\citet{kennicutt98} law are (from left to right): (1)~the flattening of the 
stellar IMF for masses below $1\,\rm M_{\sun}$ \citep{chabrier03} compared to 
the single power law of the \citet{salpeter55} IMF assumed by 
\citet{kennicutt98}; (2)~the gravitational lensing magnification factor, $\mu$, 
deduced for the lensed images A2 and A3 and for the star-forming blob from the 
gravitational lens modeling (Sect.~\ref{sect:reconstruction}); and (3)~the 
light loss through the spectrograph slit, which we estimate, from the 
convolution of the seeing profile and the slit width, assuming an average 
seeing of $0.7\arcsec$ during our NIR observations and a slit width of 
$0.9\arcsec$, to be a factor of 1.1. The H$\alpha$ luminosity (in 
$\rm erg\,s^{-1}$) can be directly derived from the H$\alpha$ line flux, 
knowing the redshift of the galaxy. The intrinsic H$\alpha$ fluxes of the 
8~o'clock arc, corrected for dust extinction $E(B-V)_{\rm gas}$ determined from 
the H$\alpha$/H$\gamma$ Balmer decrement as described in 
Sect.~\ref{sect:dustextinction}, are listed in Table~\ref{tab:fluxes}.

We obtain as an extinction/lensing-corrected star-formation rate 
$\rm SFR_{H\alpha}^{A2} = 279\pm 45\,M_{\sun}\,yr^{-1}$ for the image A2 and 
$\rm SFR_{H\alpha}^{A3} = 239\pm 99\,M_{\sun}\,yr^{-1}$ for the image A3, again
consistent within $1\,\sigma$ errors. These results also agree well with the 
previous estimates of the SFR in the 8~o'clock arc derived by \citet{allam07} 
and \citet{finkelstein09}. This confirms that this LBG has a high SFR, which 
falls toward the upper end of the $\rm SFR_{H\alpha}$ distribution of $z\sim 2$ 
LBGs \citep[e.g.,][]{erb06b}.

Regarding the line-profile decomposition of the image A2, the 
extinction/lensing-corrected SFR of the star-forming blob, 
$\rm SFR_{H\alpha}^{blob} = 33\pm 19\,M_{\sun}\,yr^{-1}$, appears to be 
significantly lower relative to the SFR of the main galaxy, 
$\rm SFR_{H\alpha}^{main} = 207\pm 80\,M_{\sun}\,yr^{-1}$. The SFR hence 
provides the first clear evidence of possibly different physical conditions 
characterizing the blob compared to those of the main galaxy.

An independent measure of the star-formation rate is provided by the UV 
continuum from OB stars. From the $V$-band photometry 
(Table~\ref{tab:photometry}) we have the rest-frame UV continuum flux, 
$f_{\nu}(1600)$, near 1600\,\AA\ derived from the definition of AB magnitudes. 
The corresponding luminosity, $L_{\nu}(1600)$ in $\rm erg\,s^{-1}\,Hz^{-1}$, in 
turn implies
\begin{equation}
{\rm SFR_{UV}\,(M_{\sun}\,yr^{-1})} = 1.4\times 10^{-28} L_{\nu}(1600)\times 
    \frac{1}{1.8}\times \frac{1}{\mu}\,,
\end{equation}
with the \citet{kennicutt98} scaling between $L_{\rm UV}$ and SFR (valid over
the wavelength range $1500-2800$\,\AA), and applying the same corrections as 
above for the \citet{chabrier03} IMF and magnification factors. The SFR so 
derived has then to be corrected for dust extinction. 

Using the stellar dust extinction $E(B-V)_{\rm stars}$ determined from the SED
modeling (Sect.~\ref{sect:SED}) because the UV continuum is dominated by the 
stellar light from OB stars, and the \citet{calzetti00} reddening curve, we
obtain as an extinction/lensing-corrected star-formation rate from the 
rest-frame UV continuum $\rm SFR_{UV}^{A2} = 156\pm 110\,M_{\sun}\,yr^{-1}$ for 
the image A2 and $\rm SFR_{UV}^{A3} = 198\pm 140\,M_{\sun}\,yr^{-1}$ for the 
image A3. These dust-corrected rest-frame UV SFRs agree very well with 
$\rm SFR_{SED} = 162^{+124}_{-95}$ derived from the full SED fit (see 
Sect.~\ref{sect:SED}), while they are lower than the $\rm SFR_{H\alpha}$ 
measurements reported above and corrected for the gas-phase (nebular) color 
excess $E(B-V)_{\rm gas}$. This difference essentially comes from the different 
dust extinction corrections applied, $E(B-V)_{\rm stars}$ versus 
$E(B-V)_{\rm gas}$. Indeed, the attenuation of the stellar UV light and the 
nebular emission may differ in general \citep{calzetti00,calzetti01}, and also 
appears to differ, although marginally, for this object, as discussed in 
Sect.~\ref{sect:dustcorrections}.

%

\subsection{H$\alpha$ and H$\beta$ equivalent widths}
\label{sect:EW}

%

The H$\beta$ and H$\alpha$ equivalent widths, $W(\rm H\beta)$ and 
$W(\rm H\alpha)$, provide an additional tool to investigate the star-formation 
history. As the ratio of the H$\beta$(H$\alpha$) luminosity to the underlying 
stellar continuum, $W(\rm H\beta)$($W(\rm H\alpha)$) is a measure of the 
current to past average star formation. With a reliable stellar continuum 
detection in the vicinity of the H$\beta$ line and a more tentative one around 
the H$\alpha$ line because of the larger noise in the X-shooter spectra at 
these wavelengths, we nevertheless obtain a direct measure of the H$\beta$ and 
H$\alpha$ rest-frame equivalent widths: $W_0(\rm H\beta)^{A2} = 30\pm 2\,\AA$ 
and $W_0(\rm H\alpha)^{A2} = 134\pm 67\,\AA$ for the image A2, and 
$W_0(\rm H\beta)^{A3} = 29\pm 3\,\AA$ and $W_0(\rm H\alpha)^{A3} = 
139\pm 72\,\AA$ for the image A3. The measurements show an excellent agreement 
between A2 and A3. No slit loss and dust extinction corrections are applied in 
this case, which indirectly assumes that the nebular emission lines and the 
stellar continuum suffer the same attenuation. In Table~\ref{tab:properties} 
we also provide the H$\beta$ and H$\alpha$ equivalent widths corrected for the 
different dust attenuations, nebular emission versus stellar continuum. We used 
these equivalent widths as an additional cross-check of the SED fits, which 
include the treatment of both the nebular emission continuum and lines (see 
Sect.~\ref{sect:SED}).

%

\subsection{Mass}
\label{sect:masses}

\subsubsection{Gas mass}
\label{sect:gasmass}

%

Because of the lack of CO measurements, one usually relies on the 
Schmidt-Kennicutt relation between star-formation rate and gas-mass surface 
density to determine the gas masses of high-redshift galaxies 
\citep[e.g.,][]{erb06b,forster09}. This relation has been established for local
star-forming galaxies \citep[e.g.,][]{kennicutt98}, and its validity has 
recently been tested at high redshifts from direct measurements of CO molecular 
lines in bright sub-millimeter galaxies \citep{bouche07,tacconi06,tacconi08} 
and in several rest-UV/optically selected star-forming galaxies (BzK and BX 
objects) at $z\sim 1-2.5$ \citep{daddi08,daddi10,genzel10,tacconi10}. All show 
that both local and high-redshift star-forming galaxies lie approximately along 
the universal Schmidt-Kennicutt relation. To estimate the gas mass in the 
8~o'clock arc, we use the \citet{bouche07} calibration valid for both local and 
high-redshift galaxies:
\begin{equation} \label{eq:gasmass}
M_{\rm gas}\,({\rm M_{\sun}) = 3.66\times 10^8\, 
     (SFR\,(M_{\sun}\,yr^{-1}}))^{0.58}\, (r_{1/2}\,(\rm kpc))^{0.83}\,.
\end{equation}
The half-light radius, $r_{1/2}$, is directly measured from the reconstructed 
source plane image, obtained from the gravitational lens modeling 
(Sect.~\ref{sect:reconstruction}). By using this half-light radius to estimate 
the gas mass, we indirectly assume that $r_{1/2}$, as measured from the 
rest-frame UV light, also applies to the molecular gas. In applying 
(\ref{eq:gasmass}), we take half of the inferred star-formation rate for the 
area enclosed within $r_{1/2}$, and multiply by two to get the total gas mass.
\citet{genzel10} have recently proposed a revised calibration of the 
Schmidt-Kennicutt relation, which they compare with previous calibrations, and 
discuss in detail the various possible origins of differences (see their 
Sect.~4.4). The authors conclude that the total systematic uncertainty of slope 
determinations probably is $\pm 0.2$ to $\pm 0.25$.

Using the star-formation rates as derived from the H$\alpha$ luminosity and
corrected for gas-phase dust extinction and lensing, we derive a gas mass 
$M_{\rm gas}^{\rm A2} = (20.9\pm 3.9)\times 10^9\,\rm M_{\sun}$ for the image 
A2 and $M_{\rm gas}^{\rm A3} = (19.1\pm 6.4)\times 10^9\,\rm M_{\sun}$ for the 
image A3. These measurements obtained for the two lensed images of the 
8~o'clock arc agree very well. They are reliable within a factor of up to 
$\sim 1.4$, depending on the star-formation rates we are referring to, 
$\rm SFR_{H\alpha}$ versus $\rm SFR_{UV}$. The overall order of magnitude of 
the inferred gas masses agrees with the gas-mass estimate of the 8~o'clock arc 
obtained by \citet{finkelstein09}.

Considering the separate contributions from the main galaxy and the 
star-forming blob identified in the image A2, we derive a significantly smaller 
gas mass for the blob, $M_{\rm gas}^{\rm blob} = (2.2\pm 0.9)\times 
10^9\,\rm M_{\sun}$, than for the main galaxy, $M_{\rm gas}^{\rm main} = 
(17.6\pm 5.6)\times 10^9\,\rm M_{\sun}$. This results from the one order of 
magnitude lower $\rm SFR_{H\alpha}$ and the three times smaller half-light 
radius of the blob. The link between the blob and the main galaxy is discussed 
in Sect.~\ref{sect:blob}.

%

\subsubsection{Dynamical mass}
\label{sect:dynmass}

%

Dynamical masses can be calculated from the line widths via the relation
\citep{erb06b}:
\begin{equation}
M_{\rm dyn}\,({\rm M_{\sun}}) = 
    \frac{C\,(\sigma\,({\rm km\,s^{-1}}))^2\,r_{1/2}\,({\rm kpc})}{G}\,,
\end{equation}
where $G = 4.3\times 10^{-6}\,\rm kpc\,(km\,s^{-1})^2\,M_{\sun}^{-1}$ is the 
gravitational constant, and the factor $C$ depends on the galaxy's mass density
profile, the velocity anisotropy, the relative contributions to $\sigma$ from 
random motions or rotation, and the assumption of a spherical or disk-like
system. Under the assumption that a disk rotation is appropriate, we begin with 
$M_{\rm dyn,rot}\,(r<r_{1/2}) = \upsilon_{\rm true}^2 r_{1/2}/G$. 
We incorporate an average inclination correction 
$\langle \upsilon_{\rm true}\rangle = 
\upsilon_{\rm FWHM}/\langle \sin(i) \rangle$, where $\langle \sin(i) \rangle = 
\pi/4$ and the observed half-width velocity $\upsilon_{\rm FWHM} = \rm FWHM/2 = 
2.355 \sigma/2$. Hence, for rotation-dominated objects the enclosed dynamical 
mass within the half-light radius, $r_{1/2}$, is $M_{\rm dyn,rot}\,(r<r_{1/2}) 
= (2.25 \sigma^2 r_{1/2})/G$. We then multiply this resulting mass by two to 
obtain the total dynamical mass. For dispersion-dominated objects, we apply the 
isotropic virial estimator with $M_{\rm dyn,disp} = (6.7 \sigma^2 r_{1/2})/G$, 
appropriate for a variety of galactic mass distributions \citep{binney08}. In 
this case, $M_{\rm dyn,disp}$ represents the total dynamical mass.

The Gaussian two-component decomposition of the line profiles within the image 
A2, proposed to be associated with the main galaxy and the star-forming blob,
respectively, as observed in the reconstructed source plane image (see
Fig.~\ref{fig:reconstruction}), yields a dynamical mass $M_{\rm dyn}^{\rm main}
= (8.4\pm 2.7)\times 10^9\,\rm M_{\sun}$ ($(12.6\pm 4.0)\times 
10^9\,\rm M_{\sun}$) for the main component of the galaxy and 
$M_{\rm dyn}^{\rm blob} = (1.1\pm 0.3)\times 10^9\,\rm M_{\sun}$ 
($(1.7\pm 0.5)\times 10^9\,\rm M_{\sun}$) for the blob, under the assumption of
rotation(dispersion)-dominated kinematics. This confirms the one order of
magnitude smaller mass of the blob with respect to the main component of the 
galaxy. 

With the dynamical masses in hand, we may try to infer whether the star-forming 
blob is in rotation around the main core of the galaxy. Knowing the distance, 
$d = 1.2\pm 0.1\,\rm kpc$, between the blob and the galaxy from the source 
reconstruction (Sect.~\ref{sect:reconstruction}), we can calculate the expected 
velocity of the blob in rotation around the galaxy from 
$\upsilon_{\rm expected}^{\rm blob-galaxy} = \sqrt{G M_{\rm dyn}^{\rm main}/d}$, 
and check whether it agrees with the observed velocity of the blob relative to 
the galaxy, $\upsilon_{\rm obs}^{\rm blob-galaxy} = 130\pm 5\,\rm km\,s^{-1}$, 
as measured from the Gaussian two-component best-fit of the line profiles 
(velocity difference between the Gaussian centroids of the blob and the galaxy; 
Table~\ref{tab:paramGaussians}). With the derived dynamical mass of the main 
core of the galaxy (rotation-dominated value), we obtain with an average 
inclination correction $\upsilon_{\rm expected}^{\rm blob-galaxy} 
\langle \sin(i) \rangle = 136\pm 23\,\rm km\,s^{-1}$, which excellently agrees 
with the observed velocity of the blob relative to the galaxy. This strongly
suggests that the blob is in rotation around the main core of the galaxy. 
Considering the dynamical mass of the main core of the galaxy as derived 
assuming a dispersion-dominated kinematics, we still observe an agreement with 
$\upsilon_{\rm obs}^{\rm blob-galaxy}$ within $1.5\,\sigma$ error. 

Assuming the blob and the main component are embedded in the same system, with 
the blob in rotation around the main core of the galaxy, then the size of the 
combined $\rm main + blob$ system should be used for the system size estimate 
and the best-fit Gaussian two-component velocity dispersions, given in 
Table~\ref{tab:paramGaussians}, summed up in quadrature, should be used for the 
system half-width velocity estimate. As a result, we find a total dynamical 
mass $M_{\rm dyn}^{\rm A2} = (16.0\pm 4.9)\times 10^9\,\rm M_{\sun}$ for the 
image A2 and $M_{\rm dyn}^{\rm A3} = (20.2\pm 15.9)\times 10^9\,\rm M_{\sun}$ 
for the image A3. Again the two measurements obtained for the two lensed 
images A2 and A3 of the 8~o'clock arc give consistent values within $1\,\sigma$ 
errors, as expected. On the other hand, if the blob and the main component of 
the galaxy are merging, then the dynamical masses of each should be combined 
for a total dynamical mass. This yields a total dynamical mass $M_{\rm dyn} = 
(14.3\pm 4.5)\times 10^9\,\rm M_{\sun}$.

%

\section{Summary of the results and discussion}
\label{sect:discussion}

%

\subsection{Extinction correction and star formation}
\label{sect:dustcorrections}

Thanks to the gravitational lensing, the 8~o'clock arc offers a rare 
opportunity to compare the dust extinction corrections as derived from the 
Balmer decrement and from the SED modeling of multi-band photometric data in an 
LBG at $z=2.7350$. Studies of local starburst galaxies show that the ionized 
gas is more attenuated than stars \citep{calzetti00,calzetti01}, a difference 
that is usually interpreted as indicating that young hot ionizing stars are 
associated with dustier regions than the bulk of the (cooler) stellar 
population across the galaxies. \citet{calzetti01} gives the following relation 
between the color excess observed for gas and stars: $E(B-V)_{\rm stars} = 
0.44\times E(B-V)_{\rm gas}$. In the 8~o'clock arc, we find for gas 
$E(B-V)^{\rm A2,A3}_{\rm gas} = 0.30\pm 0.07$ and for stars $E(B-V)_{\rm stars} 
= 0.19\pm 0.04$. Taking these results at face value, we observe a trend in the
8~o'clock arc toward a larger dust attenuation in the ionized gas than in 
stars, in line with the \citet{calzetti01} relation, although this may appear 
marginal because the two color excesses are comparable within $2-3\,\sigma$ 
when considering the measurement and SED modeling uncertainties. The UV slope, 
$\beta$, measured from the observed $(V-I)$ colors is $\beta^{\rm A2,A3} = 
-1.35\pm 0.43$, typical of $z\sim 2.5$ LBGs at the same UV luminosities 
($M_{\rm UV} = -22.3$ after correction for lensing). This translates to 
$E(B-V)^{\rm A2,A3}_{\beta} = 0.19\pm 0.10$, which excellently agrees with the
stellar dust extinction $E(B-V)_{\rm stars}$ derived from the SED fits. The
larger uncertainty on $E(B-V)_{\beta}$, nevertheless, reduces the confidence
level on the observed difference between gaseous and stellar dust attenuations.

Differences between gaseous and stellar extinction corrections have recently 
been investigated by \citet{yoshikawa10} at $z\sim 2$ in $K$-band selected 
star-forming galaxies. The authors' comparison of the H$\alpha$, UV, and 
Spitzer/MIPS 24\,$\mu \rm m$ fluxes shows that the SFRs of lower SFR galaxies 
($\lesssim 100\,\rm M_{\sun}\,yr^{-1}$) agree well for the equal-extinction 
case $E({\rm B-V})_{\rm stars} = E({\rm B-V})_{\rm gas}$, while those of higher 
SFR galaxies agree better for the \citet{calzetti01} relation, although 
$\rm SFR_{H\alpha}$ are systematically higher than $\rm SFR_{UV}$ by 0.3\,dex 
\citep[see also][]{hayashi09}. This suggests that the relation between dust 
properties of stellar continuum and nebular lines is different depending on the 
intrinsic SFR. The 8~o'clock arc with $\rm SFR_{H\alpha,\rm UV} > 
100\,M_{\sun}\,yr^{-1}$, as well as another lensed LBG the Cosmic Horseshoe at 
$z=2.38$ \citep{hainline09}, tend to support this trend. However, whether these 
differences are true or caused by other effect(s) needs to be established more 
firmly.

The stellar masses derived from the SED modeling are often compared with the
SFRs to search for correlation. While some authors report a significant 
correlation, others find that a fraction of galaxies deviates from the bulk of 
the distribution, showing a general trend of increasing SFRs with stellar 
masses \citep{erb06c,daddi07,daddi08,daddi10,hayashi09,magdis10,yoshikawa10}. 
\citet{erb06b} found that these outliers are galaxies with 
$M_{\rm dyn}/M_{\rm stars} > 10$, young ages $< 100\,\rm Myr$, high H$\alpha$ 
rest-frame equivalent widths $W_0(\rm H\alpha) > 200$\,\AA, and high gas 
fractions $\mu_{\rm gas} > 60$\%. \citet{yoshikawa10} further confirmed 
that whatever dust extinction correction is applied (\citet{calzetti01}-type or 
equal-type), these outliers remain outliers, and are characterized by 
particularly high specific star-formation rates, 
${\rm SSFR = SFR}/M_{\rm stars} > 10\,\rm Gyr^{-1}$. They thus also support 
young ages for these galaxies, with the majority of their stellar mass being 
formed in a recent starburst.

%

\begin{figure}[!]
\centering
\includegraphics[width=9cm,clip]{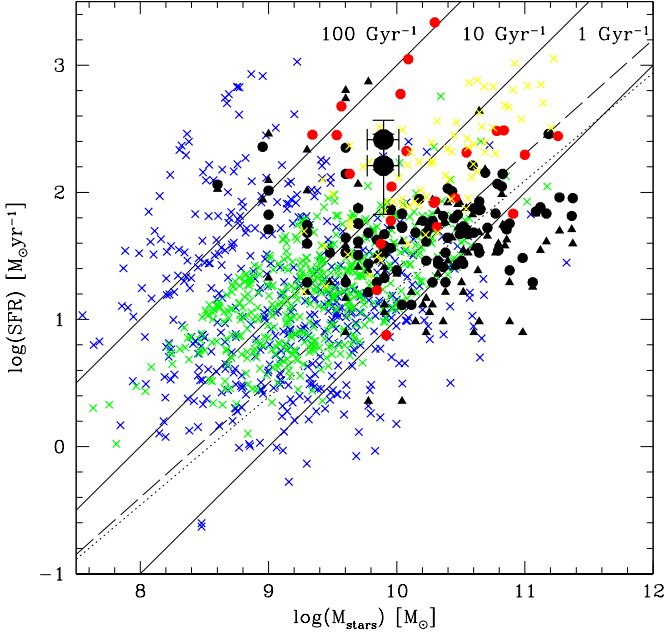}
\caption{Comparison of the derived extinction-corrected SFRs and stellar mass 
of the 8~o'clock arc (large black circles, showing the results with/without 
nebular emission and with $\rm SFR_{H\alpha}$/$\rm SFR_{SED}$) with those 
from other samples of LBGs at $z\sim 2-3$. The black triangles and small 
circles show the $z\sim 2$ objects from \citet{erb06b,erb06c} with 
dust-corrected SFRs determined from H$\alpha$ and from SED fitting, 
respectively. The red circles show the objects from \citet{yoshikawa10} with 
$\rm SFR_{H\alpha}$ corrected for dust extinction by assuming a different 
nebular and stellar attenuation. The yellow crosses are the $z\sim 3$ IRAC 
detected LBGs from \citet{magdis10}. The green and blue crosses correspond to 
the $U$-dropout ($z\sim 3$) sample from the GOODS South field, analyzed by 
\citet{debarros11}, with different assumptions in the SED fits (green crosses: 
constant SFR only, no nebular emission; blue crosses: variable star-formation 
histories, including nebular emission). All results were rescaled to the 
\citet{chabrier03} IMF, if necessary. The dotted and dashed lines show the 
mass-SFR relations suggested by \citet{sawicki07} and \citet{daddi07} for 
$z\sim 2$ galaxies. The thin solid lines show the locii of constant specific
star-formation rates, $\rm SSFR = 100$, 10, and $1\,\rm Gyr^{-1}$, 
respectively.}
\label{fig:mass-SFR}
\end{figure}
%

In Fig.~\ref{fig:mass-SFR} we plot the extinction-corrected star-formation 
rates and stellar masses of the 8~o'clock arc and other comparison samples of 
star-forming galaxies at $z\sim 2-3$ (LBGs, $K$-band selected star-forming 
galaxies, and IRAC-detected LBGs from \citet{erb06b,erb06c}, 
\citet{yoshikawa10}, \citet{magdis10}, and \citet{debarros11}). The results 
were scaled to the \citet{chabrier03} IMF, if necessary. All plotted stellar 
masses were obtained from SED fits, but the detailed models that were used 
differ somehow, which may introduce some scatter between the different studies. 
The 8~o'clock arc shows a relatively high SFR for its stellar mass, 
corresponding to a specific SFR on the order of $\rm SSFR^{A2,A3} = 
33\pm 19\,Gyr^{-1}$. Compared to the two samples with $\rm SFR_{H\alpha}$ 
measurements, we see that this is clearly higher than the typical values 
derived by \citet{erb06c}, although not exceptionally higher, and when compared 
to the sample of \citet{yoshikawa10}. We do not know if there is a physical 
reason for this high SSFR, but we note that the 8~o'clock arc also shows a 
relatively high gas fraction (see below) and a young age, features pointed out 
by \citet{erb06c} for their objects that have the largest specific 
star-formation rates.

%

\subsection{Gas fraction}

As an important result, we assess at the same time the stellar mass, the gas
mass, and the dynamical mass of the 8~o'clock arc. Those masses were obtained 
from independent quantities, the multi-wavelength photometry, the H$\alpha$ 
luminosity, and the nebular emission line width, respectively. The sum of the 
stellar mass and gas mass determines the baryonic mass $M_{\rm bar} = 
M_{\rm stars} + M_{\rm gas}$, and the fraction $\mu_{\rm gas} = 
M_{\rm gas}/(M_{\rm stars} + M_{\rm gas})$ determines the gas fraction. We
derive for the 8~o'clock arc the baryonic mass $M^{\rm A2,A3}_{\rm bar} = 
(27.9\pm 7.5)\times 10^9\,\rm M_{\sun}$, which can be compared to the dynamical 
mass $M^{\rm A2,A3}_{\rm dyn} = (18.1\pm 10.4)\times 10^9\,\rm M_{\sun}$ 
($(14.3\pm 4.5)\times 10^9\,\rm M_{\sun}$) as derived under the 
rotation(dispersion)-dominated kinematics assumption (see 
Table~\ref{tab:properties} and Sect.~\ref{sect:dynmass}). The two measurements 
agree within a factor of $1.5-2$, which is very good given the significant 
uncertainties carried by the respective masses (observational plus systematic), 
and is in line with the $M_{\rm bar}$ versus $M_{\rm dyn}$ dispersion observed 
for $z\sim 2$ LBGs \citep{erb06b}. We find that the 8~o'clock arc has a high 
gas fraction with $\mu_{\rm gas} \simeq 72$\%, contrary to the 
\citet{finkelstein09} result, where the authors inferred a $\mu_{\rm gas}$ of 
only $\sim 12$\% because of their large stellar mass estimate. A high gas 
fraction is yet the overall trend of $z\sim 2$ LBGs that show a mean gas 
fraction of 50\%. Moreover, the derived physical properties in our lensed LBG 
support the observed correlations of decreasing $\mu_{\rm gas}$ with increasing 
stellar mass and age \citep{erb06b,reddy06}. The 8~o'clock arc consequently 
appears as a young starburst with a still significant gas fraction and a low 
fraction of baryonic mass that already turned out into stellar mass.

%

\subsection{A fundamental mass, SFR, and metallicity relation beyond 
$z\simeq 2.5$\,?}

Recently, \citet{mannucci10} studied the dependence of gas-phase metallicity,
$12+\log (\rm O/H)$, on stellar mass, $M_{\rm stars}$, and star-formation rate, 
$\rm SFR_{H\alpha}$, and found a fundamental mass, SFR, and metallicity 
relation satisfied by local SDSS galaxies' metallicities down to a dispersion 
of about 0.05\,dex. The well-known mass-metallicity relation is in fact one 
particular projection of this fundamental relation into one plane, and the 
observed evolution of the mass-metallicity relation \citep{savaglio05,erb06a,
maiolino08,mannucci09} is caused by the increase of the average SFR with 
redshift, which results in sampling different parts of the fundamental relation 
at different redshifts. This fundamental relation seems to hold up to 
$z\simeq 2.5$ without any evolution, which means that the same physical 
processes are in place in the local Universe and at high redshifts. Beyond this 
redshift, the few available measurements \citep{maiolino08,mannucci09} show 
hints of evolution.

The 8~o'clock arc at $z=2.7350$ provides a nice opportunity to test the
fundamental relation in the high-redshift regime thanks to the available
accurate measurements of metallicity, stellar mass, and SFR. Considering the 
parameterization of the metallicity as a linear combination of 
$\rm SFR_{H\alpha}$ and $M_{\rm stars}$ (relation (5) in \citet{mannucci10}), 
we expect for the 8~o'clock arc a metallicity $12+\log(\rm O/H) \simeq 8.49$, 
which agrees very well with the measured metallicity, 
$12+\log(\rm O/H)^{A2,A3}_{N2} = 8.41\pm 0.19$, derived from the N2 index (see 
Table~\ref{tab:properties}). This suggests that the fundamental relation may 
hold up beyond $z=2.5$, as also supported by two other lensed LBGs at 
$2.5 < z < 3.5$ studied by \citet{richard11}. Larger statistics is clearly 
needed to determine the trend in the fundamental relation at higher redshifts. 

%

\subsection{The blob and its interpretation}
\label{sect:blob}

Spatially-resolved kinematics of massive $z=1.5-2.5$ star-forming galaxies from
the SINS survey \citep{genzel06,genzel08,forster09} provide some of the most 
convincing evidence for the existence of large, turbulent, gas-rich rotating 
disks and for secular processes in non-major merging systems playing a 
significant role in growing galaxies at $z\sim 2$. Key properties of these 
early disks are high intrinsic velocity dispersions of $\sigma \sim 
30-80\,\rm km\,s^{-1}$, and high gas fractions $\mu_{\rm gas} > 30$\%. In 
several of these disks, kpc-sized star-forming clumps are observed in H$\alpha$ 
emission. Clump stellar and gas masses are $10^8-10^{9.5}\,\rm M_{\sun}$, and 
akin to the ubiquitous rest-frame UV clumps in HST optical imaging of $z>1$ 
star-forming galaxies \citep[e.g.,][]{elmegreen09}. Clumpy galaxies are, 
indeed, not rare. These giant massive clumps could be the telltale signature of 
disk fragmentation through Jeans instabilities, and could migrate inward to the 
disk center and coalesce into a slowly rotating bulge \citep{bournaud07,
elmegreen08}. However, this bulge formation scenario is still a matter of 
current debate because it hinges on the survival of clumps in the presence of 
stellar feedback (outflows driven by stellar winds, supernovae and radiation 
pressure). State-of-art theory and simulations \citep{krumholz10,murray10,
genel10} lead to contradictory predictions on clump stability, and urgently 
call for empirical constraints. 

With a flux accounting for about 14\% of the integrated H$\alpha$ flux of the 
entire galaxy (lensing-corrected), a mass one order of magnitude lower than the 
main core of the galaxy, a half-light radius smaller than that of all 
$z\sim 1-2$ galaxies studied so far \citep[e.g.,][]{law07,law09,forster09}, and 
a small spatial offset from the main core of the galaxy $d = 
1.2\pm 0.1\,\rm kpc$, we believe that the resolved star-forming blob in the 
8~o'clock arc galaxy is one of these clumps commonly observed in high-redshift 
objects. The physical properties of the blob, such as its gas and dynamical 
masses $M_{\rm gas,dyn} = (1.1-2.2)\times 10^9\,\rm M_{\sun}$, star-formation 
rate $\rm SFR_{H\alpha} = 33\pm 19\,M_{\sun}\,yr^{-1}$, and size $r_{1/2} = 
0.53\pm 0.05\,\rm kpc$ (see Table~\ref{tab:properties}), indeed much resemble 
the properties of the clumps studied by \citet{swinbank09}, \citet{jones10}, 
and \citet{genzel11}. In particular, the blob perfectly satisfies the 
mass-size and size-SFR relations observed for high-redshift clumps. It hence 
additionally supports the conjecture that these high-redshift \ion{H}{ii} 
regions are comparable in size and mass to the largest local star-forming 
complexes, and are consistent with the mass-size relation observed locally, but 
have a $\sim 100$~$\times$ higher SFR than in local spiral galaxies. The blob 
in the 8~o'clock arc even nicely extrapolates the size-SFR relation toward a 
domain with higher SFRs and sizes. Moreover, similarly to the modest 
clump-to-clump and inside-out disk variations observed in the oxygen abundances 
by \citet{genzel11}, we do also see some changes between the metallicity of the 
main galaxy and the blob, with the blob showing a trend toward a lower 
metallicity by about 0.17\,dex. According to a simple model 
(Sect.~\ref{sect:dynmass}), the blob appears to be in rotation around the main 
core of the galaxy, which suggests that the 8~o'clock arc LBG may be one of 
these turbulent, gas-rich rotating disk. Additional knowledge of detailed 
physical properties of these clumps is a very useful input to models trying to 
predict their formation and evolution within high-redshift star-forming 
galaxies.

%

\begin{acknowledgements}
We are very grateful to Fr\'ed\'eric Courbin and Roser Pell\'o for helpful
discussions, and we thank the anonymous referee for her/his very careful and 
constructive report. We thank the Paranal ESO Observatory for the high-quality 
data acquired in service mode for this programme. M.D.-Z. and D.S. acknowledge 
support from the Swiss National Science Foundation, and J.R. from an EU 
Marie-Curie Fellowship.
\end{acknowledgements}

%

\end{document}